\documentclass[
 aps,           
pra,            
twocolumn,      
letterpaper,    
showpacs,       
preprintnumbers,
amsmath,        
amssymb,        
floatfix]{revtex4-2}

\usepackage[T1]{fontenc}
\usepackage{graphicx}    
\usepackage{color}      
\usepackage{bm}         
\usepackage{braket}

\begin{document}
\title{\textbf{Quantum discord of Ganssian states in an expanding universe}} 

 \author{Siwei Li}
 \author{Xiaofen Huang} 
 \email{huangxf1206@163.com}
 \affiliation{School of Mathematics and Statistics, Hainan Normal University, Haikou 571158, China}
 
 \begin{abstract}
We investigate the redistribution of continuous-variable quantum discord within the framework of an expanding universe. We find that quantum discord exhibits stronger sensitivity to the spacetime expansion rate than to the expansion volume. As both the expansion rate and expansion volume increase, the initial quantum discord shared by the two bosonic modes decays, while quantum discord is induced in additional mode pairs by the underlying spacetime expansion, signaling a global redistribution of quantum correlations across the system. Specifically, the induced discord is largest for cross-observer bosonic-antibosonic pairs, followed by same-observer bosonic-antibosonic pairs, and smallest for the pair of antibosonic modes. Furthermore, our quantum discord analysis demonstrates that particles with lower momentum and optimal mass serve as more favorable candidates for extracting information about the expanding universe. This work substantially enriches the theoretical framework of quantum discord in expanding spacetimes, and provides new perspectives as well as a solid theoretical foundation for further investigations.
\end{abstract}

\maketitle

\section{Introduction}

Quantum discord is a physical quantity that characterizes quantum correlations based on measurements. It holds a fundamental position in quantum information science and has a very wide range of applications \cite{luo2008using,li2008classical,radhakrishnan2020multipartite,inui2020entanglement,zhou2020quantum}.
Unlike entanglement, quantum discord can describe a broader class of nonclassical correlations: it takes a nonzero value even for separable states, and generally exhibits stronger robustness against decoherence than entanglement.
Analytical solutions for this measure have been derived for families of two-qubit states with special structures \cite{geometric_measure_quantum_discord,quantum_discord_two_qubit_systems,quantum_discord_geometry_bell_diagonal,quantum_discord_geometry_class_two_qubit,quantum_discord_two_qubit_analytical_small_error,quantum_discord_x_states_one_variable,analytical_formula_quantum_discord_two_qubit_x,super_quantum_discord_two_qubit_x,computing_quantum_discord_np_complete}.
For multipartite quantum systems, researchers have proposed generalized forms including the squared quantum discord and multiqubit geometric discord \cite{exploring_multipartite_quantum_correlations_square_discord,quantum_discord_multiqubit_systems,geometric_discord_multiqubit_systems,monogamy_deficit_quantum_correlations_multipartite,conditions_monogamy_quantum_correlations_ghz_w}.
For continuous-variable states, Giorda \textit{et al.} presented a closed-form analytical formulation for squeezed thermal states (STS) \cite{giorda_paris_2010}.

Recent work in relativistic quantum information has focused on non-inertial spacetimes, with extensive studies investigating the behavior of quantum correlations, including entanglement \cite{zhang2025entanglement,liu2025entanglement,liu2023fermionic,wang2020genuine,mi2024impact,Li2023Quantum}, coherence \cite{li2025multiqubit,wu2021quantum,liao2025quantum,Kaczmarek2025Coherence}, nonlocality \cite{mi2025genuine,zhang2023hawking,Kaczmarek2024Signatures,Kaczmarek2026Nonlocal}, uncertainty relations \cite{Wang2024Entropic},  quantum steering \cite{wu2025fermionic,wu2025gaussian,liu2018influence,Mi2026Gaussian}, quantum discord \cite{huang2026,non_markovian_effect_quantum_discord,sudden_change_quantum_discord_single_qubit_noise,quantum_discord_resource_remote_state_preparation,observing_operational_significance_discord_consumption,quantum_discord_bounds_distributed_entanglement,quantum_coherence_geometric_quantum_discord,converting_coherence_quantum_correlations,quantum_coherence_multipartite_systems,unified_view_quantum_correlations_quantum_coherence,relative_quantum_coherence_incompatibility_quantum_correlations_states} and so on. Key results show these correlations are degraded by information loss due to Hawking radiation \cite{liu2018satellite,liu2019influence,wang2011multipartite,torres2019entanglement,wu2020quantum}, findings that advance both our understanding of quantum information in curved backgrounds and the study of the black hole information paradox and entanglement entropy.
Despite numerous significant advances in related studies, the dynamical evolution of continuous-variable quantum discord in an expanding universe remains poorly understood to date.

In this work, we study the redistribution of continuous-variable quantum discord of a scalar field in an expanding universe, where Alice and Bob initially share a two-mode squeezed Gaussian state in the asymptotic past. As established in previous works, the cosmic expansion drives the asymptotic-past vacuum state into a thermal state in the asymptotic future \cite{Ball2006PLA,MartinMartinez2012CQG,Liu2020QIP,Fuentes2010PRD}, which can be equivalently described as a Gaussian channel acting on the initial Gaussian state in the quantum information framework. The two-mode squeezed Gaussian state is chosen for two reasons: it is a canonical continuous-variable entangled state approximating EPR pairs with arbitrary precision, and it is experimentally accessible for existing continuous-variable quantum information setups \cite{Einstein1935EPR,Braunstein2005RMP}.

The rest of this paper is organized as follows. Sec. II is devoted to a review of the Gaussian channel picture of cosmic expansion and the formal framework of continuous-variable quantum discord. In Sec. III, we systematically investigate the dynamical characteristics of quantum discord under the cosmic expansion background, and present the main findings on its evolutionary behavior. At last, Sec. IV gives a brief summary of the whole work.

\section{Description of Cosmic Expansion via Gaussian Channels and Quantum Discord of Gaussian States}

\subsection{Expansion of the universe described by Gaussian channels}

We commence our analysis in a $1 + 1$-dimensional Robertson-Walker expanding spacetime, whose metric
takes the form \cite{Ball2006PLA,Fuentes2010PRD} 
\begin{equation}
	ds^2 = dt^2 - \left[a(t)\right]^2\, dx^2,
	\label{eq:rw_1p1_metric}
\end{equation}
where $a(t)$ denotes the  scale factor. Introducing the conformal time $\eta$, related to the cosmological  time $t$ by
\begin{equation}
	\eta = \int_0^t \frac{d\tau}{a(\tau)},
	\label{eq:conformal_time_def}
\end{equation}
one may rewrite the Robertson--Walker expanding universe can be rewritten as \cite{Ball2006PLA,MartinMartinez2012CQG,Liu2020QIP,Fuentes2010PRD}
\begin{equation}
	ds^2 = \left[a(\eta)\right]^2 \left(d\eta^2 - dx^2\right).
	\label{eq:rw_conformal_metric}
\end{equation}

Within this framework, the conformal scale factor reads
\begin{equation}
	\left[a(\eta)\right]^2 = 1 + \epsilon\left(1 + \tanh(\upsilon\eta)\right),
	\label{eq:scale_factor_tanh}
\end{equation}
where the  parameters $\epsilon$ and $\upsilon$ describe the volume and  rate of the cosmic expansion, respectively. One can readily verify that the spacetime geometry reduces to flat Minkowski spacetime in both the asymptotic past and asymptotic future: in the limit $\eta \to -\infty$, the metric simplifies to $ds^2 = d\eta^2 - dx^2$, while for $\eta \to +\infty$ it takes the form $ds^2 = (1 + 2\epsilon)\left(d\eta^2 - dx^2\right)$. As a result, a global timelike Killing vector field exists in both limits, ensuring that the particle spectrum of the quantum field is well-defined within these two limits.

We next consider a real scalar field $\phi(x,\eta)$ propagating in the expanding Robertson--Walker spacetime, whose dynamics is governed by the Klein--Gordon equation
\begin{equation}
	\left(\frac{1}{\sqrt{|g|}} \partial_\mu \sqrt{|g|}\, g^{\mu\nu} \partial_\nu + m^2\right) \Phi = 0.
	\label{eq:kg_covariant}
\end{equation}
Upon solving the Klein--Gordon equation in the two limits $\eta \to \pm\infty$, we construct the orthonormal basis of $u^\mathrm{in}$ modes valid over the distant past (``in'') domain and its complementary counterpart $u^\mathrm{out}$ mode basis for the far future (``out'') domain. Making use of the inner product, the Bogoliubov transformation relating the modes $u_k^\mathrm{in}$ and $u_k^\mathrm{out}$ can be expressed as
\begin{equation}
	u_k^\mathrm{in}(x,\eta) = \alpha_k u_k^\mathrm{out}(x,\eta) + \beta_k u_{-k}^{\mathrm{out}*}(x,\eta),
	\label{eq:bogoliubov_mode}
\end{equation}
where the Bogoliubov coefficients take the explicit form
\begin{align}
	\alpha_k &= \sqrt{\frac{\omega_\mathrm{out}}{\omega_\mathrm{in}}} \,
	\frac{\Gamma\left(1 - \frac{\mathrm{i}\omega_\mathrm{in}}{\upsilon}\right)
		\Gamma\left(-\frac{\mathrm{i}\omega_\mathrm{out}}{\upsilon}\right)}
	{\Gamma\left(1 - \frac{\mathrm{i}\omega_+}{\upsilon}\right)
		\Gamma\left(-\frac{\mathrm{i}\omega_+}{\upsilon}\right)}, \\
	\beta_k &= \sqrt{\frac{\omega_\mathrm{out}}{\omega_\mathrm{in}}} \,
	\frac{\Gamma\left(1 - \frac{\mathrm{i}\omega_\mathrm{in}}{\upsilon}\right)
		\Gamma\left(\frac{\mathrm{i}\omega_\mathrm{out}}{\upsilon}\right)}
	{\Gamma\left(1 + \frac{\mathrm{i}\omega_-}{\upsilon}\right)
		\Gamma\left(\frac{\mathrm{i}\omega_-}{\upsilon}\right)},
\end{align}
where $\Gamma$ denotes the gamma function, $\omega_\mathrm{in} = \sqrt{k^2 + m^2}$, $\omega_\mathrm{out} = \sqrt{k^2 + m^2(1+2\epsilon)}$, and $\omega_\pm = \frac{1}{2}\left(\omega_\mathrm{out} \pm \omega_\mathrm{in}\right)$. Elementary algebraic manipulation verifies that the Bogoliubov coefficients obey $|\alpha_k|^2 - |\beta_k|^2 = 1$. To streamline subsequent derivations, we introduce
\begin{equation}
	\theta_k^2 = \left|\frac{\beta_k}{\alpha_k}\right|^2
	= \frac{\sinh^2\left(\pi \frac{\omega_-}{\upsilon}\right)}
	{\sinh^2\left(\pi \frac{\omega_+}{\upsilon}\right)},
	\label{eq:theta_k_def}
\end{equation}
from which we immediately deduce
\begin{equation}
	|\alpha_k|^2 = \frac{1}{1-\theta_k^2}, \quad |\beta_k|^2 = \frac{\theta_k^2}{1-\theta_k^2},
	\label{eq:bogoliubov_mod_sq}
\end{equation}
where $|\beta_k|^2$ corresponds to the mean particle occupation number generated in the ``out'' mode $k$. Accordingly, the limit $\theta_k^2 \to 0$ implies the average particle count of mode $k$ vanishes, whereas $\theta_k^2 \to 1$ signifies that the mean particle number for mode $k$ diverges to infinity.

The bosonic annihilation and creation operators obey the Bogoliubov linear mapping relations
\begin{align}
	b_{\mathrm{in},k} &= \alpha_k^* b_{\mathrm{out},k} - \beta_k^* b^\dagger_{\mathrm{out},-k},
	\label{eq:bogoliubov_op_annih} \\
	b^\dagger_{\mathrm{in},k} &= \alpha_k b^\dagger_{\mathrm{out},k} - \beta_k b_{\mathrm{out},-k},
	\label{eq:bogoliubov_op_create}
\end{align}
where $b_{\mathrm{in},k}$ and $b^\dagger_{\mathrm{in},k}$ are the bosonic annihilation and creation operators acting on the states in the asymptotic past, $b_{\mathrm{out},k}$ and $b^\dagger_{\mathrm{out},k}$ are the bosonic annihilation and creation operators acting on the states in the asymptotic future, and $b_{\mathrm{out},-k}$ and $b^\dagger_{\mathrm{out},-k}$ act as the annihilation and creation operators for antiparticles, respectively. We use $b_{\mathrm{in},k} \left.|0_k\rangle\right|_\mathrm{in} = 0$ to find the relation between the ``in'' vacuum state and the ``out'' vacuum state. Substituting $b_{\mathrm{in},k}$ via Eq.~\eqref{eq:bogoliubov_op_annih}, we obtain
\begin{equation}
	\left( \alpha_k^* b_{\mathrm{out},k} - \beta_k^* b^\dagger_{\mathrm{out},-k} \right) |0_k\rangle_{\mathrm{in}} = 0.
	\label{eq:in_vacuum_condition}
\end{equation}

By virtue of the Bogoliubov normalization constraint, the ``in'' vacuum state can be written in the asymptotic future as
\begin{equation}
	|0_k\rangle_{\mathrm{in}} = \sum_{n=0}^{\infty} A_n |n_k\rangle_{\mathrm{out}} |n_{-k}\rangle_{\mathrm{out}},
	\label{eq:in_vacuum_expansion}
\end{equation}
where $A_n = \sqrt{1-\theta_k^2} \left( \frac{\beta_k^*}{\alpha_k^*} \right)^n$, $n_k$ represents the boson number, and $n_{-k}$ represents antiboson number. This decomposition demonstrates that the primordial vacuum state $|0_k\rangle_{\mathrm{in}}$ evolves into a canonical two-mode squeezed state within the far future. Upon rotating the squeezing parameter and absorbing the global phase factor, we arrive at \cite{Caves1985a,Caves1985b,Adesso2007PRA,Adesso2012CQG}
\begin{equation}
	\begin{split}
	|0_k\rangle_{\mathrm{in}} &= \sqrt{1-\theta_k^2} \sum_{n=0}^{\infty} \theta_k^n |n_k\rangle_{\mathrm{out}} |n_{-k}\rangle_{\mathrm{out}} \\
	&= U_k |0_k\rangle |0_{-k}\rangle,
	\label{eq:in_vacuum_squeezed}
	\end{split}
\end{equation}
were $U_k = \exp\left[ r_k \left( b^\dagger_{\mathrm{out},k} b^\dagger_{\mathrm{out},-k} - b_{\mathrm{out},k} b_{\mathrm{out},-k} \right) \right]$ stands for a two-mode squeezing operator, where the squeezing parameter $r_k$ is fixed by the algebraic relation $\cosh(r_k) = |\alpha_k|$. It is imperative to emphasize that $U_k$ implements a Gaussian quantum operation, which preserves the Gaussian statistical profile of all input quantum states. Hence, Eq.~\eqref{eq:in_vacuum_squeezed} demonstrates that the cosmological expansion of the Robertson--Walker spacetime can be modeled as a Gaussian channel equivalent to a bosonic amplification process. Within the phase-space formulation of quantum mechanics, the action induced by $U_k$ corresponds to a symplectic transformation characterized by the matrix
\begin{equation}
	S_k = \frac{1}{\sqrt{1-\theta_k^2}}
	\begin{pmatrix}
		I_2 & \theta_k Z_2 \\
		\theta_k Z_2 & I_2
	\end{pmatrix},
	\label{eq:symplectic_squeezing}
\end{equation}
where $I_2$ denotes the unity matrix in $2\times2$ space, and $Z_2$ denotes the third Pauli matrix.

\subsection{Quantum discord of Gaussian states}

Within the present subsection, we survey core definitions and symbolic conventions relevant to Gaussian states. We focus on an $n$-mode continuous-variable system equipped with state space $\mathcal{H}=\mathcal{H}_1\otimes \mathcal{H}_2 \otimes \dots \otimes \mathcal{H}_n$, where each subspace $\mathcal{H}_i$ satisfying $1 \leq i \leq n$ corresponds to an infinite-dimensional complex Hilbert space. For each $m=1,2,\dots,n$, we label the single-mode Fock basis as the set $\{|j_m\rangle\}_{j_m=0}^{\infty} \subset \mathcal{H}_m$.

Denote by $\mathcal{S(H)}$  the set of all quantum states (that is, positive bounded linear operators with trace 1) on $\mathcal{H}$.
For any state $\rho\in \mathcal{S(H)}$, its characteristic function $\chi_{\rho}$ can be expressed as
\begin{equation}
	\chi_{\rho}(z)=\rm{Tr} \rho W(z),
\end{equation}
where $z=(x_1, y_1, x_2, y_2, \dots, x_n, y_n)^t$, $W(z)={\rm{exp}}({\mathrm{i}}R^t z)$ is the Weyl displacement operator and $R=(\hat{R}_1, \hat{R}_2, \dots, \hat{R}_{2n})=(\hat{X}_1, \hat{Y}_1, \dots, \hat{X}_n, \hat{Y}_n) $.  Here, as usual, $\hat{X}_p=\hat{a}_p+\hat{a}_p^{\dagger}$, and  
$\hat{Y}_p=\mathrm{i}(\hat{a}_p-\hat{a}_p^{\dagger})$ ($p=1, 2, \dots, n$) respectively stand for the position and momentum operators,
with  $\hat{a}_p^{\dagger}$ and 
$\hat{a}_p$  the creation and annihilation
operators in the $p$-th  mode $\mathcal{H}_p$ satisfying the canonical commutation relation
\[
[\hat{a}_p, \hat{a}_q^{\dagger}]=\delta_{pq}I, ~~[\hat{a}_p^{\dagger}, \hat{a}_q^{\dagger}]=[\hat{a}_p, \hat{a}_q]=0,
\]
where $ p, q=1, 2, \dots, n$.

Assume that $\rho$ has finite second moment. The displacement vector (or mean) $\bar{d}$ of $\rho$  is given by
\begin{equation}
	\bar{d}=(\langle \hat{R}_1 \rangle, \dots, \langle \hat{R}_{2n}\rangle)^t
	=(\mathrm{Tr}(\rho \hat{R}_1), \dots, \mathrm{Tr}(\rho \hat{R}_{2n}))     
\end{equation}
and the covariance matrix $\sigma=(\sigma_{pq})$ of $\rho $ is defined as
\begin{equation}
	\sigma_{pq}=\frac{1}{2}\langle  \Delta \hat{R}_p \Delta \hat{R}_q+\Delta \hat{R}_q\Delta \hat{R}_p\rangle,   
\end{equation}
where $\Delta \hat{R}_p=\hat{R}_p-\langle \hat{R}_p \rangle$ \cite{Braunstein2005QICV}. 
It should be noted that a covariance matrix $\sigma$ is symmetric and must satisfy the uncertainty principle
\[
\sigma+\mathrm{i}\Omega_n\geq 0,
\]
where $\Omega_n=\oplus_{i=1}^n\Omega_i$ with 
$\Omega_i=\begin{pmatrix} 0 & 1\\-1 & 0    \end{pmatrix}$ for each $i$ \cite{Simon1994QNM}. 
In addition, $\sigma \ge 0$ as $\sigma +\mathrm{i}\Omega_n \ge 0$.

Every property possessed by a Gaussian state is uniquely fixed by the first and second statistical moments associated with the quadrature operators. Local unitary operations enable arbitrary tuning of the first moments without altering any informationally meaningful features, such as entropy and various correlation quantifiers. Given this invariance property, one may safely fix all first moments to zero; accordingly, the second-order moments constitute the only necessary descriptors for Gaussian states.

Based on our research, we initially consider a two-mode Gaussian state $\rho$ shared by Alice and Bob, which is 
specified by its covaricance matrix, a real,  symmetric and 
positive matrix:
\begin{equation}
	\sigma = \begin{pmatrix}
		\mathcal{A} & \mathcal{C} \\
		\mathcal{C}^\mathrm{T} & \mathcal{B}
	\end{pmatrix},
	\label{eq:cov_sts}
\end{equation}
where $\mathcal{A} = \mathrm{diag}(a,a)$, $\mathcal{B} = \mathrm{diag}(b,b)$, and $\mathcal{C} = \mathrm{diag}(c_{1},c_{2})$ are $2\times2$ diagonal submatrices.

The quantum discord of a Gaussian state can be expressed as \cite{giorda_paris_2010}
\begin{equation}
	D(\sigma) = h\left(\sqrt{\lambda_2}\right) - h(d_-) - h(d_+) + \inf_{\sigma_M} h\left(\sqrt{\sigma_P}\right).
	\label{eq:gaussian_discordyiban}
\end{equation}
Here, $
	\sigma_P = \mathcal{A} - \mathcal{C} \left( \mathcal{B} + \sigma_M \right)^{-1} \mathcal{C}^\mathrm{T},
	\label{eq:schur_complement}
$
where $
	\sigma_M = 
	\begin{pmatrix}
		\alpha & \gamma \\
		\gamma & \beta
	\end{pmatrix},
$
with fixed parameters $\alpha, \beta \in \mathbb{R}^+$ and $\gamma \in \mathbb{R}$.
 The  function takes the form
\begin{equation}
	h(x) = \left(x+\frac{1}{2}\right)\ln\left(x+\frac{1}{2}\right) - \left(x-\frac{1}{2}\right)\ln\left(x-\frac{1}{2}\right),
	\label{eq:h_function}
\end{equation}
as well as the quantities $d_\pm$ satisfying
\begin{equation}
	d_\pm^2 = \frac{1}{2}\left( \Delta \pm \sqrt{\Delta^2 - 4\lambda_4} \right),
	\Delta = \lambda_1 + \lambda_2 + 2\lambda_3,
	\label{eq:d_pm_def}
\end{equation}
where $\lambda_1 = \det\mathcal{A}$, $\lambda_2 = \det\mathcal{B}$,  $\lambda_3 = \det\mathcal{C}$, and  $\lambda_4 = \det\sigma$.

For the generic bipartite squeezed thermal states (STS), the matrix elements of the covariance matrix \eqref{eq:cov_sts} are given by
\begin{equation}
	\begin{split}
		&a = \frac{1}{2} + N_\xi + N_1(1+N_\xi) + N_2 N_\xi, \\
		&b = \frac{1}{2} + N_\xi + N_2(1+N_\xi) + N_1 N_\xi, \\
		&c_1 = -c_2 = \left(1 + N_1 + N_2\right)\sqrt{N_\xi(1+N_\xi)},
	\end{split}\notag
	\label{eq:cov_elements_sts}
\end{equation}
where $N_\xi=\sinh^2 \xi$ and $N_j$ is the mean 
thermal photon number of mode $j$, and $j\in \{1, 2\}$.

The quantum discord of the generic bipartite squeezed thermal states can be expressed as \cite{giorda_paris_2010}
\begin{equation}
	\begin{split}
		D(\sigma) &= h(\sqrt{\lambda_2}) - h(d_-) - h(d_+) \\
		&\quad + h\left( \frac{\sqrt{\lambda_1} + 2\sqrt{\lambda_1 \lambda_2} + 2\lambda_3}{1 + 2\sqrt{\lambda_2}} \right).
	\end{split}
	\label{eq:quantum_discord_sts}
\end{equation}

\section{The Effect of the Expanding Universe on Gaussian Quantum Discord}
In this section, we begin by considering a pure two-mode squeezed Gaussian state shared between Alice and Bob in the asymptotic past. The corresponding covariance matrix reads \cite{adesso_fuentes_schuller_ericsson_2007}
\begin{equation}
	\sigma^\mathrm{in}_{AB} =
	\begin{pmatrix}
		\dfrac{\cosh(2s)}{2}\, I_2 & \dfrac{\sinh(2s)}{2}\, Z_2 \\[1.2em]
		\dfrac{\sinh(2s)}{2}\, Z_2 & \dfrac{\cosh(2s)}{2}\, I_2
	\end{pmatrix},
	\label{eq:tmss_covariance}
\end{equation}
where $s$ denotes the squeezing parameter.
As the cosmological expansion acts on the state via the symplectic transformation given in Eq.~\eqref{eq:symplectic_squeezing}, the initial two-mode squeezed state $\sigma^\mathrm{in}_{AB}$ prepared in the asymptotic past evolves into a four-mode Gaussian state in the asymptotic future. In other words, the Robertson--Walker spacetime expansion maps the initial two-mode squeezed state of the asymptotic past onto a four-mode Gaussian state in the asymptotic future. Therefore, a complete description of the quantum system involves four modes: the bosonic mode $A$ described by Alice; the bosonic mode $B$ described by Bob; the antibosonic mode $\bar{A}$ described by anti-Alice; the antibosonic mode $\bar{B}$ described by anti-Bob. The covariance matrix describing the complete system thus reads \cite{adesso_fuentes_schuller_ericsson_2007}
\begin{equation}
	\sigma^\mathrm{out}_{AB\bar{A}\bar{B}} = \left[ S_{A,\bar{A}} \oplus S_{B,\bar{B}} \right] \left[ \sigma^\mathrm{in}_{AB} \oplus I_{\bar{A}\bar{B}} \right] \left[ S_{A,\bar{A}} \oplus S_{B,\bar{B}} \right]^\mathrm{T},
	\label{eq:4mode_cov_evolution}
\end{equation}
where $S_{A,\bar{A}}$ and $S_{B,\bar{B}}$ given by Eq.~\eqref{eq:symplectic_squeezing} are the phase-space representations of the two-mode squeezing operation, $I_{\bar{A}\bar{B}}$ denotes the $4\times 4$ identity matrix, and $\mathrm{T}$ denotes transpose.

Because Alice and Bob cannot detect the antibosonic modes, we take the partial trace over modes $\bar{A}$ and $\bar{B}$. Performing this operation on Eq.~\eqref{eq:4mode_cov_evolution}, we obtain the reduced state between modes $A$ and $B$:

\begin{equation}
	\sigma^\mathrm{out}_{AB} = \begin{pmatrix}
		\mathcal{A}_{AB} & \mathcal{C}_{AB} \\
		\mathcal{C}_{AB}^\mathrm{T} & \mathcal{B}_{AB}
	\end{pmatrix},
	\label{eq:reduced_cov_ab}
\end{equation}
where
\begin{equation}
	\begin{split}
		\mathcal{A}_{AB} &= 	\left[\frac{\cosh(2s) + \theta_k^2}{2-2\theta_k^2}\right] I_2, \\
		\mathcal{C}_{AB} &= \frac{\sinh(2s)}{2-2\theta_k^2} Z_2, \\
		\mathcal{B}_{AB} &= 	\left[\frac{\cosh(2s) + \theta_k^2}{2-2\theta_k^2}\right] I_2.\notag
	\end{split}
\end{equation}

By employing the covariance matrix given in \eqref{eq:reduced_cov_ab} together with Eq.~\eqref{eq:quantum_discord_sts}, it is feasible to evaluate the influence of the expanding universe on the quantum discord between bosonic modes.

\begin{figure*}
	\centering
	\begin{minipage}[b]{0.32\textwidth}
		\centering
		\includegraphics[width=\linewidth]{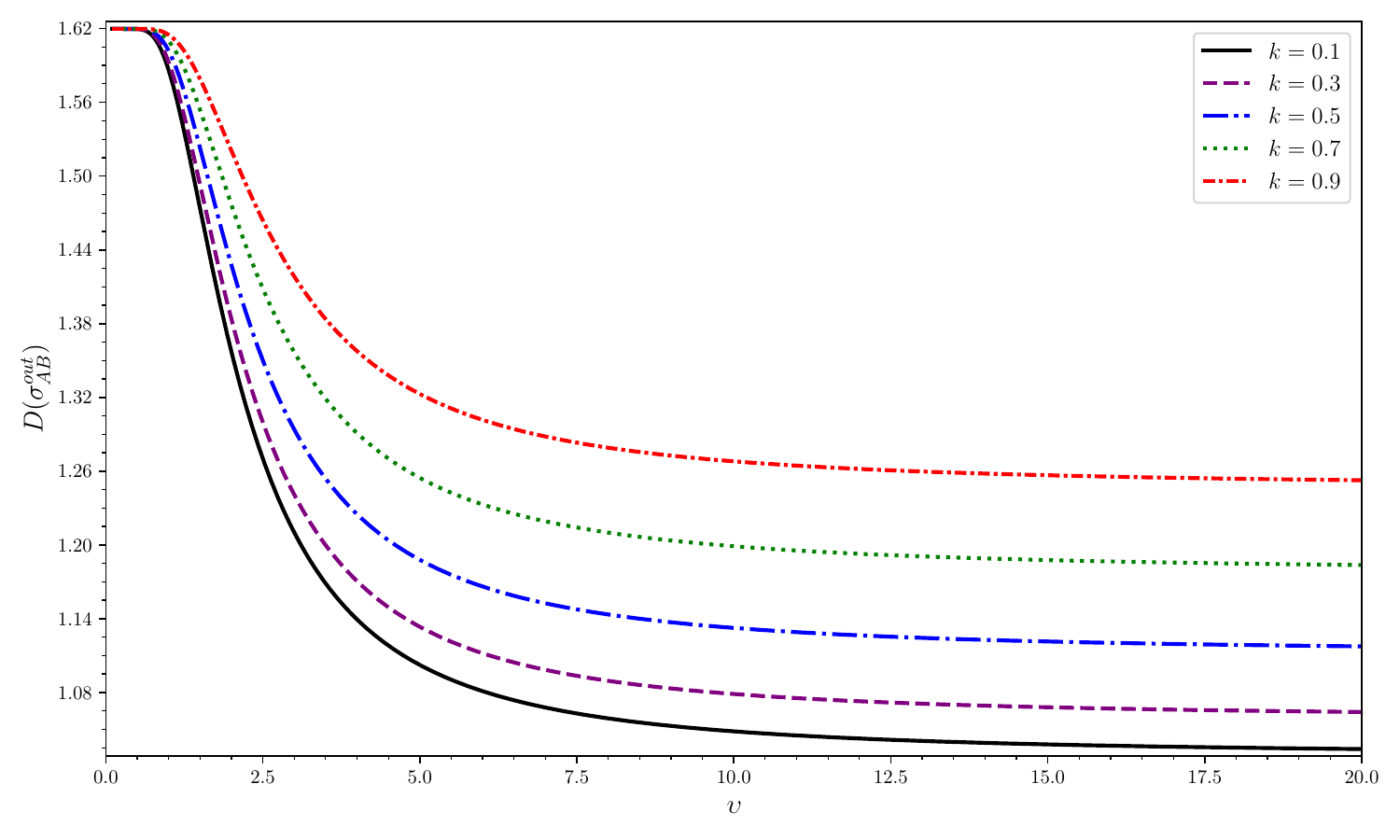}
		\\[-0.5em]
		\hspace*{1em}(a)
	\end{minipage}
	\hfill
	\begin{minipage}[b]{0.32\textwidth}
		\centering
		\includegraphics[width=\linewidth]{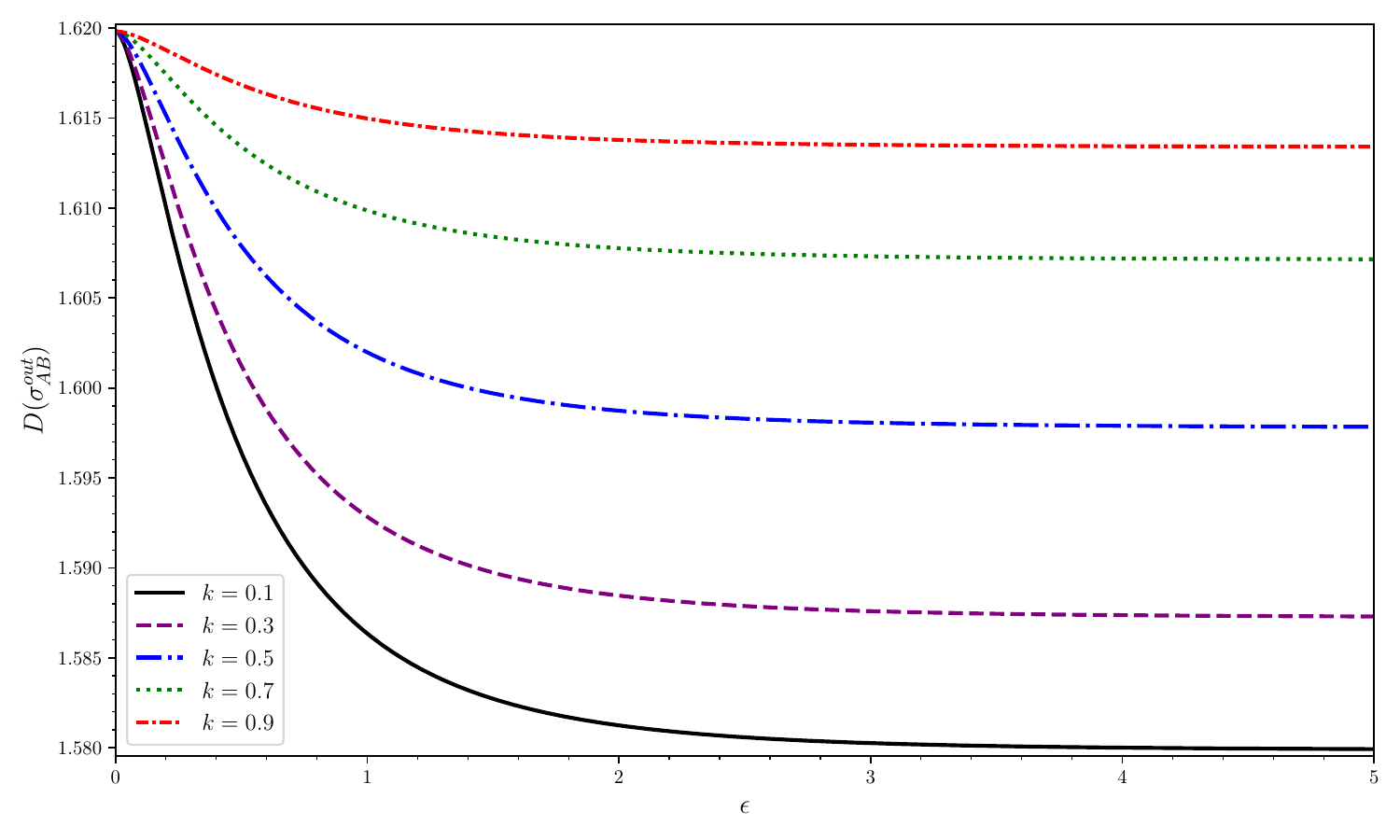}
		\\[-0.5em]
		\hspace*{1.5em}(b)
	\end{minipage}
	\hfill
	\begin{minipage}[b]{0.32\textwidth}
		\centering
		\includegraphics[width=\linewidth]{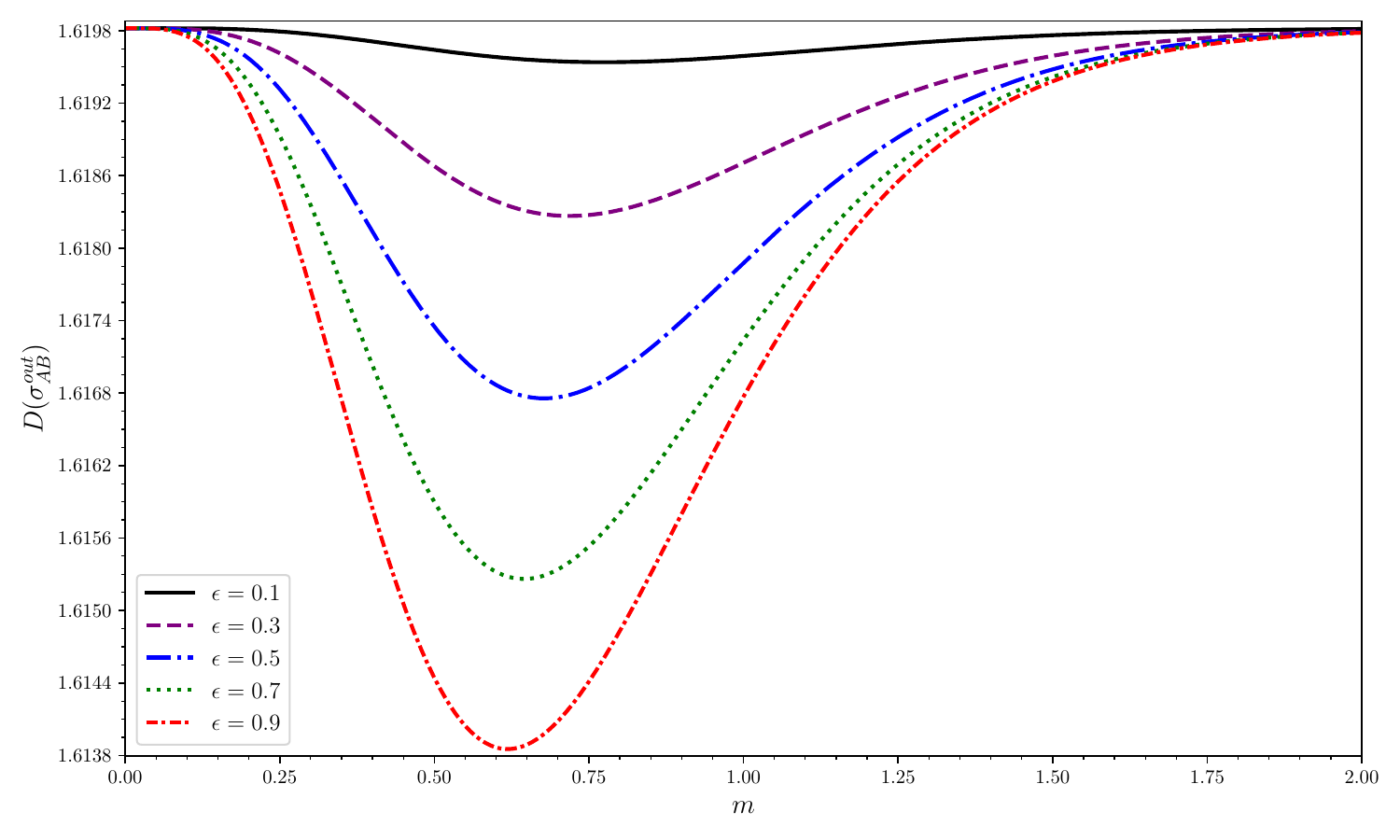}
		\\[-0.5em]
		\hspace*{1.4em}(c)
	\end{minipage}
	\caption{Quantum discord $D(\sigma^\mathrm{out}_{AB})$ between modes $A$ and $B$ as a function of the expansion rate $\upsilon$, the expansion volume $\epsilon$, and the mass $m$, respectively,  for different momenta $k$, with fixed parameters $m=\epsilon=s=1$.  }
	\label{fig:1}
\end{figure*}

In Fig.~\ref{fig:1}(a), we plot the Gaussian discord $D(\sigma^\mathrm{out}_{AB})$ between Alice and Bob as a function of the expansion rate $\upsilon$ for different momenta $k$.
Fig.~\ref{fig:1}(b), we show the Gaussian discord $D(\sigma^\mathrm{out}_{AB})$ as a function of the expansion volume $\epsilon$ for different momenta $k$.
Fig.~\ref{fig:1}(c) shows how the mass $m$ influences the Gaussian discord $D(\sigma^\mathrm{out}_{AB})$ for different expansion volumes $\epsilon$. From Fig.~\ref{fig:1}(a), we observe that the quantum discord $D(\sigma^\mathrm{out}_{AB})$ decreases as the expansion rate $\upsilon$ increases, and increases as the momentum $k$ increases. We further find that with the growth of $\upsilon$, the quantum discord converges to an asymptotic value dependent on the momentum $k$. In Fig.~\ref{fig:1}(b), the quantum discord $D(\sigma^\mathrm{out}_{AB})$ follows the same variation trend with the expansion volume $\epsilon$ and momentum $k$ as that in panel (a). In Fig.~\ref{fig:1}(c), we observe that the quantum discord $D(\sigma^\mathrm{out}_{AB})$ decreases as the expansion volume $\epsilon$ increases. We further find that as the mass $m$ increases, the quantum discord between modes $A$ and $B$ first decreases to a minimum and then increases back to its initial value. Consequently, the quantum discord of the bosonic field with a larger mass $m$ is insensitive to the expansion volume $\epsilon$. This implies that choosing bosons with an appropriate mass makes the quantum discord more sensitive to cosmological parameters.

\begin{figure}
	\centering
	\includegraphics[width=1\linewidth]{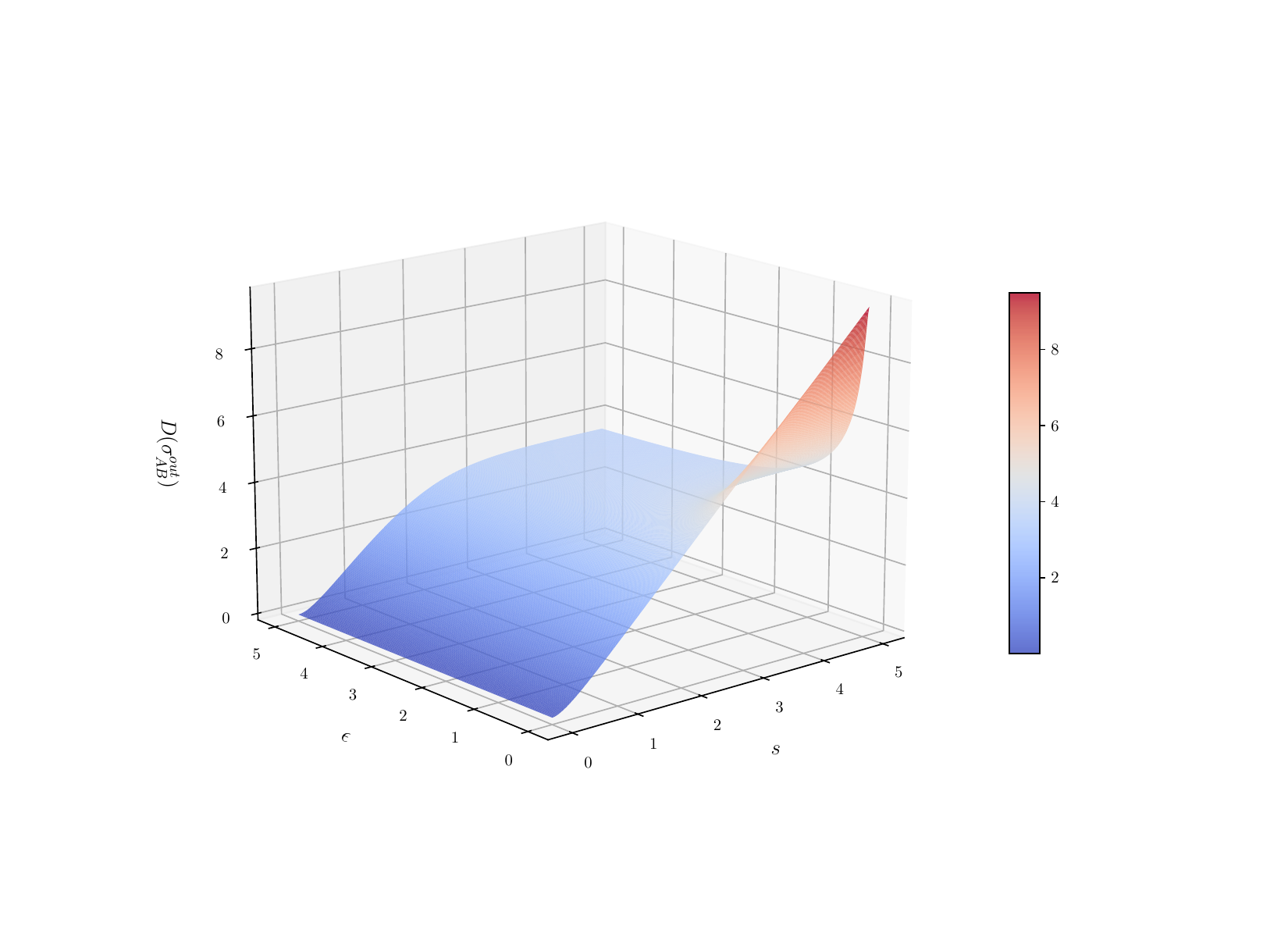}
	\caption{Quantum discord $D(\sigma^\mathrm{out}_{AB})$ between modes $A$ and $B$ as a function of the expansion volume $\epsilon$ and the squeezing parameter $s$, with fixed parameters $\frac{k}{2} = \frac{m}{2} = \upsilon = 1$.
	}
	\label{fig:2}
\end{figure}

In Fig.~\ref{fig:2}, we plot the quantum discord $D(\sigma^\mathrm{out}_{AB})$ between modes $A$ and $B$ as a function of the expansion volume $\epsilon$ and the squeezing parameter $s$. It is interesting to find that:
(i) For different values of the squeezing parameter $s$, the quantum discord $D(\sigma^\mathrm{out}_{AB})$ decreases with increasing expansion volume $\epsilon$, which is consistent with the behavior observed in Fig.~\ref{fig:1}(b).
(ii) When the initial squeezing parameter $s$ is sufficiently large, the expansion volume $\epsilon$ has a pronounced effect on the quantum discord $D(\sigma^\mathrm{out}_{AB})$, causing it to decay significantly. In contrast, when $s$ is extremely small, the quantum discord exhibits negligible variation. This indicates that the cosmological parameters can exert a significant influence on the quantum discord $D(\sigma^\mathrm{out}_{AB})$ only when the initial state is endowed with abundant quantum resources.

\begin{figure}
	\centering
	\includegraphics[width=1\linewidth]{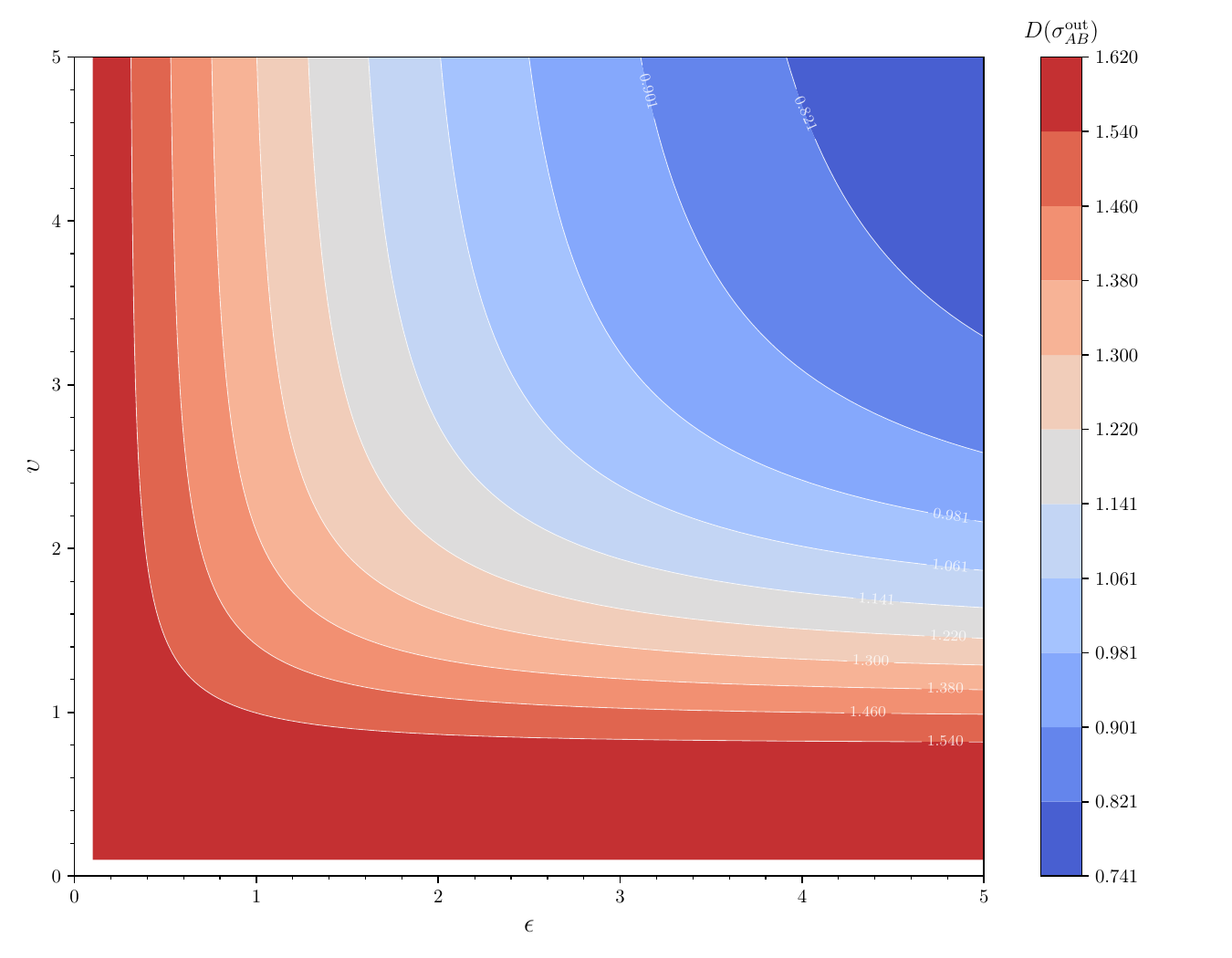}
	\caption{Quantum discord $D(\sigma^\mathrm{out}_{AB})$ between modes $A$ and $B$ as a function of the expansion volume $\epsilon$ and the expansion rate $\upsilon$, with fixed parameters $\frac{k}{2} = \frac{m}{2} = s = 1$.
	}
	\label{fig:3}
\end{figure}

In Fig.~\ref{fig:3}, we present the quantum discord $D(\sigma^\mathrm{out}_{AB})$ between modes $A$ and $B$ as a function of the expansion volume $\epsilon$ and the expansion rate $\upsilon$. We find that the expansion rate $\upsilon$ and the expansion volume $\epsilon$ exert mutually constraining effects on the quantum discord. Specifically, when the expansion rate $\upsilon$ takes a small value, even a large expansion volume $\epsilon$ exerts a negligible effect on the quantum discord. Similarly, when the expansion volume $\epsilon$ is small, even a large expansion rate $\upsilon$ also has an insignificant impact on the quantum discord. We further observe that the quantum discord $D(\sigma^\mathrm{out}_{AB})$ is more sensitive to the expansion rate $\upsilon$ than to the expansion volume $\epsilon$.

Next, we calculate the quantum discord for all remaining possible bipartitions in the four-mode quantum system, to explore the distribution of Gaussian quantum discord in an expanding spacetime. First, by tracing out the modes in subsystems $A$ and $B$, we obtain the covariance matrix $\sigma^\mathrm{out}_{\bar{A}\bar{B}}$ between modes $\bar{A}$ and $\bar{B}$.
\begin{equation}
	\sigma^\mathrm{out}_{\bar{A}\bar{B}} = \begin{pmatrix}
		\mathcal{A}_{\bar{A}\bar{B}} & \mathcal{C}_{\bar{A}\bar{B}} \\
		\mathcal{C}_{\bar{A}\bar{B}}^\mathrm{T} & \mathcal{B}_{\bar{A}\bar{B}}
	\end{pmatrix},
	\label{eq:sigma_AbarBbar_out}
\end{equation}
where
\begin{equation}
	\begin{split}
		\mathcal{A}_{\bar{A}\bar{B}} &=\left[\dfrac{\theta_k^2 \cosh(2s) + 1}{2 - 2\theta_k^2}\right]\, I_2, \\
		\mathcal{C}_{\bar{A}\bar{B}} &= \dfrac{\theta_k^2 \sinh(2s)}{2 - 2\theta_k^2}\, Z_2, \\
		\mathcal{B}_{\bar{A}\bar{B}} &=\left[\dfrac{\theta_k^2 \cosh(2s) + 1}{2 - 2\theta_k^2}\right]\, I_2.\notag
	\end{split}
\end{equation}

Secondly, by tracing out modes $B$ and $\bar{A}$, and modes $A$ and $\bar{B}$, respectively, we obtain the covariance matrices $\sigma^\mathrm{out}_{A\bar{B}}$ (Alice-anti-Bob) and $\sigma^\mathrm{out}_{\bar{A}B}$ (anti-Alice-Bob).
\begin{equation}
	\sigma^\mathrm{out}_{A\bar{B}} = \sigma^\mathrm{out}_{\bar{A}B} = \begin{pmatrix}
		\mathcal{A}_{\bar{A}B} & \mathcal{C}_{\bar{A}B} \\
		\mathcal{C}_{\bar{A}B}^\mathrm{T} & \mathcal{B}_{\bar{A}B}
	\end{pmatrix},
	\label{eq:sigma_ABbar_out}
\end{equation}
where
\begin{equation}
	\begin{split}
		\mathcal{A}_{\bar{A}B} &=\left[\dfrac{\cosh(2s) + \theta_k^2}{2 - 2\theta_k^2}\right]\, I_2, \\
		\mathcal{C}_{\bar{A}B} &= \dfrac{\theta_k \sinh(2s)}{2 - 2\theta_k^2}\, I_2, \\
		\mathcal{B}_{\bar{A}B} &=\left[\dfrac{\theta_k^2 \cosh(2s) + 1}{2 - 2\theta_k^2}\right]\, I_2.\notag
	\end{split}
\end{equation}

Finally, we turn our attention to the quantum discord between modes $A$ and $\bar{A}$, and between modes $B$ and $\bar{B}$. By tracing out modes $B$ and $\bar{B}$ and modes $A$ and $\bar{A}$, respectively, we obtain the covariance matrices for Alice-anti-Alice and Bob-anti-Bob, i.e., $\sigma^\mathrm{out}_{A\bar{A}}$ and $\sigma^\mathrm{out}_{B\bar{B}}$.
\begin{equation}
	\sigma^\mathrm{out}_{A\bar{A}} = \sigma^\mathrm{out}_{B\bar{B}} = \begin{pmatrix}
		\mathcal{A}_{B\bar{B}} & \mathcal{C}_{B\bar{B}} \\
		\mathcal{C}_{B\bar{B}}^\mathrm{T} & \mathcal{B}_{B\bar{B}}
	\end{pmatrix},
	\label{eq:sigma_AAbar_out}
\end{equation}
where
\begin{equation}
	\begin{split}
		\mathcal{A}_{B\bar{B}} &=\left[\dfrac{\cosh(2s) + \theta_k^2}{2 - 2\theta_k^2}\right]\, I_2, \\
		\mathcal{C}_{B\bar{B}} &= \dfrac{2\theta_k \cosh^2(s)}{2 - 2\theta_k^2}\, Z_2, \\
		\mathcal{B}_{B\bar{B}} &=\left[\dfrac{\theta_k^2 \cosh(2s) + 1}{2 - 2\theta_k^2}\right]\, I_2.\notag
	\end{split}
\end{equation}

Similarly, by combining the above covariance matrices \eqref{eq:sigma_AbarBbar_out}--\eqref{eq:sigma_AAbar_out} with Eqs.~\eqref{eq:gaussian_discordyiban} and \eqref{eq:quantum_discord_sts}, we can analyze the influence of the expanding universe on the quantum discord among these mode pairs.

\begin{figure*}
	\centering
	\begin{minipage}[b]{0.32\textwidth}
		\centering
		\includegraphics[width=\linewidth]{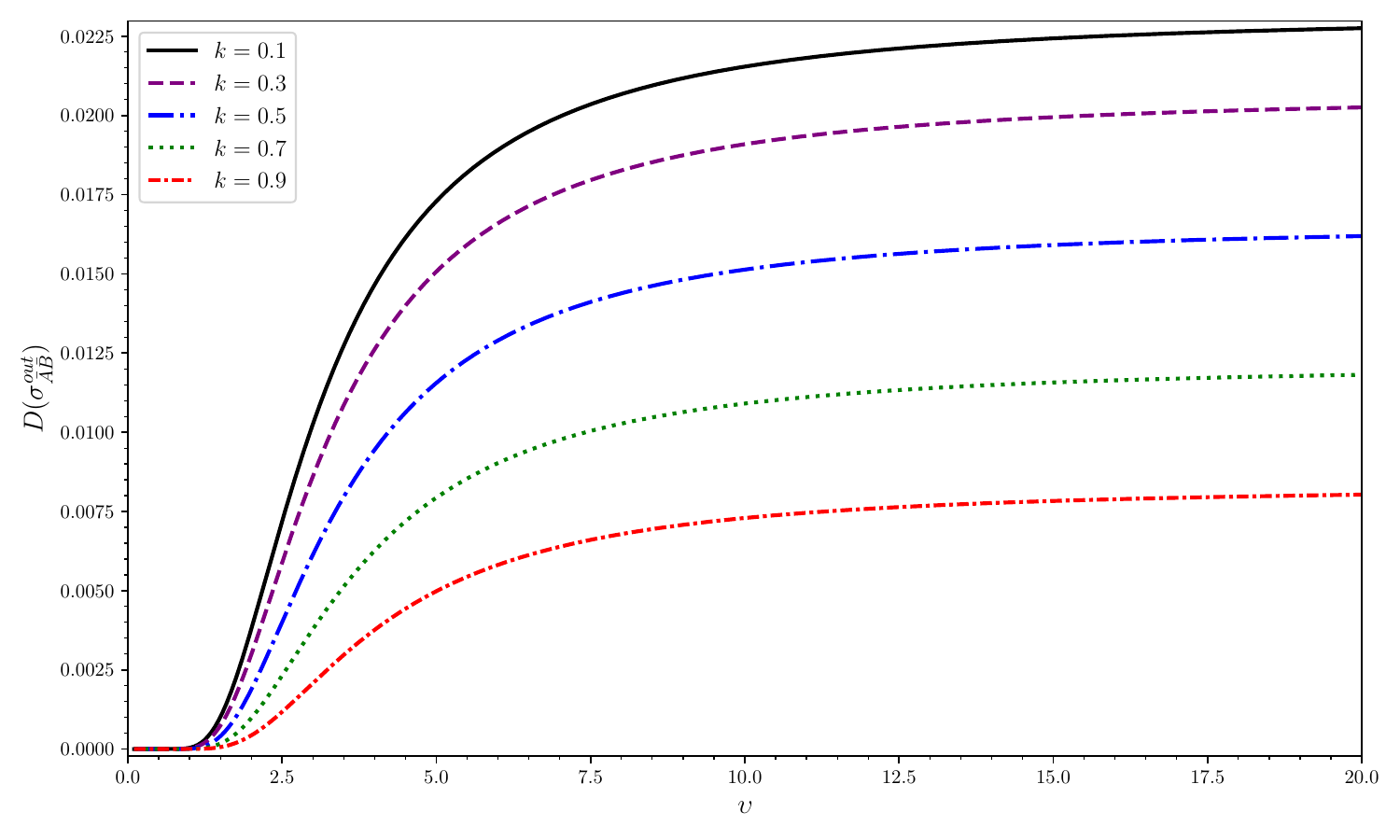}
		\\[-0.5em]
		\hspace*{1.5em}(a)
	\end{minipage}
	\hfill
	\begin{minipage}[b]{0.32\textwidth}
		\centering
		\includegraphics[width=\linewidth]{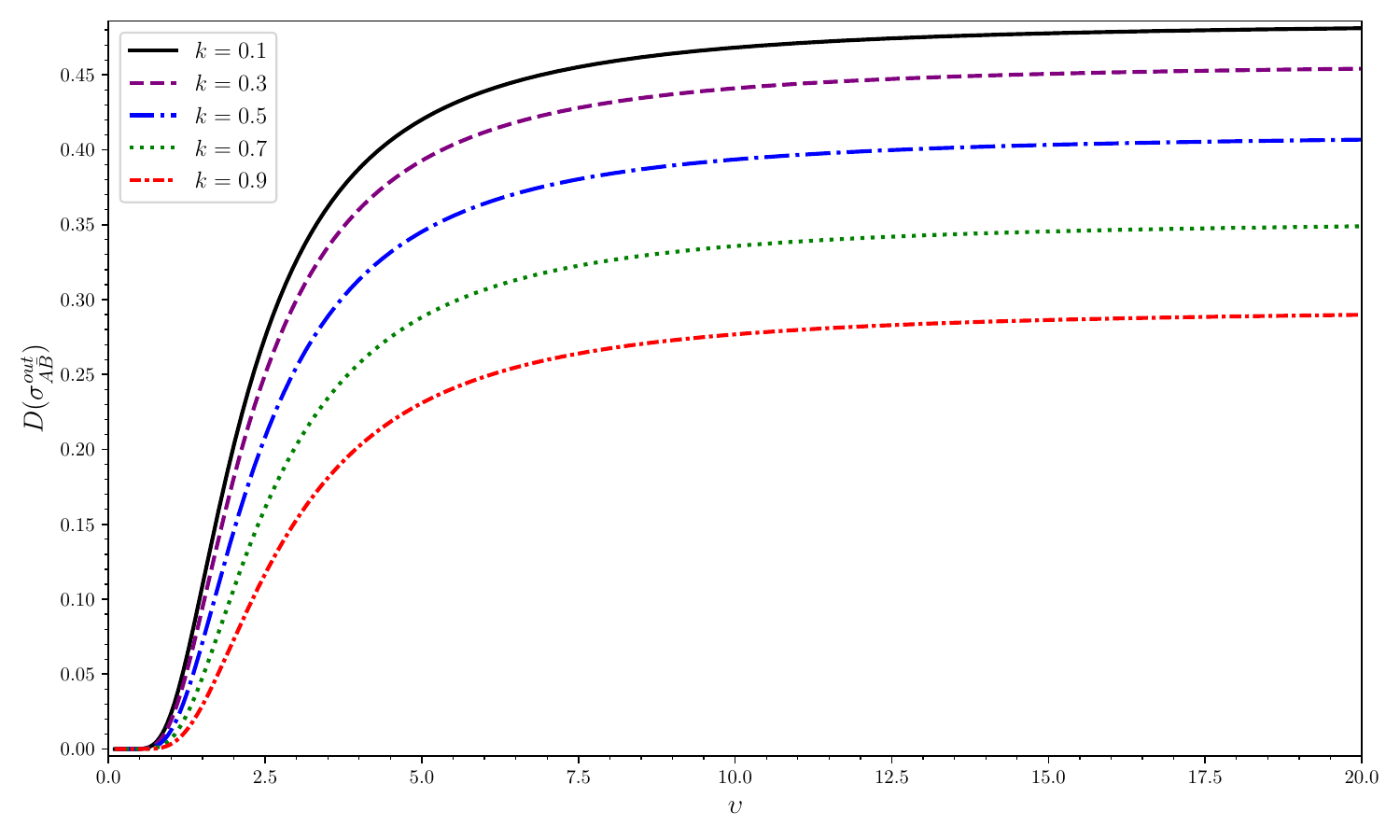}
		\\[-0.5em]
		\hspace*{1.2em}(b)
	\end{minipage}
	\hfill
	\begin{minipage}[b]{0.32\textwidth}
		\centering
		\includegraphics[width=\linewidth]{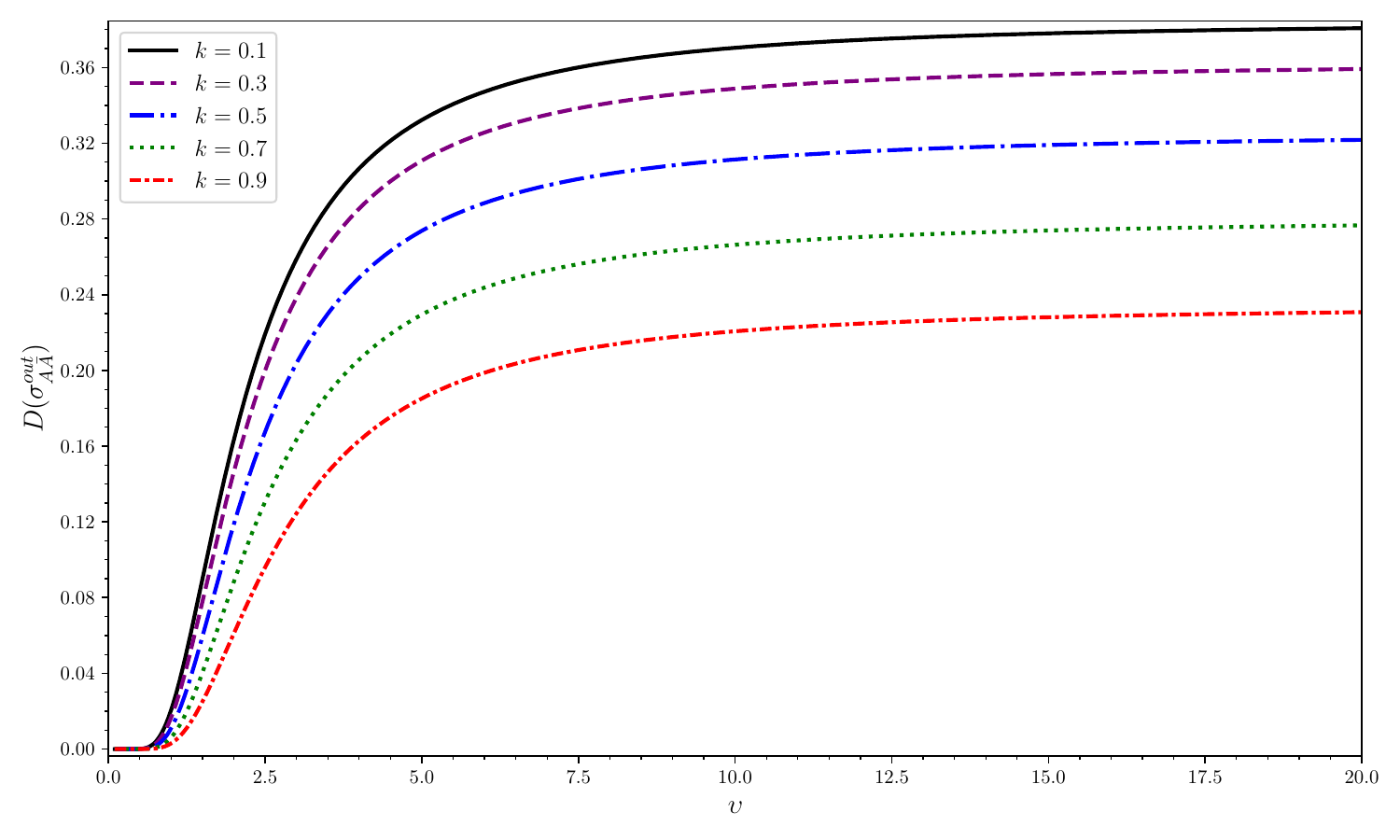}
		\\[-0.5em]
		\hspace*{1.2em}(c)
	\end{minipage}
	\caption{Panels (a), (b), and (c) show the quantum discord $D(\sigma^\mathrm{out}_{\bar{A}\bar{B}})$, $D(\sigma^\mathrm{out}_{A\bar{B}})$, and $D(\sigma^\mathrm{out}_{A\bar{A}})$ for the $\bar{A}$-$\bar{B}$, $A$-$\bar{B}$, and $A$-$\bar{A}$ mode pairs, respectively, as a function of the expansion rate $\upsilon$ for different momenta $k$, with fixed parameters $m=\epsilon=s=1$.}
	\label{fig:4}
\end{figure*}

\begin{figure*}
	\centering
	\begin{minipage}[b]{0.32\textwidth}
		\centering
		\includegraphics[width=\linewidth]{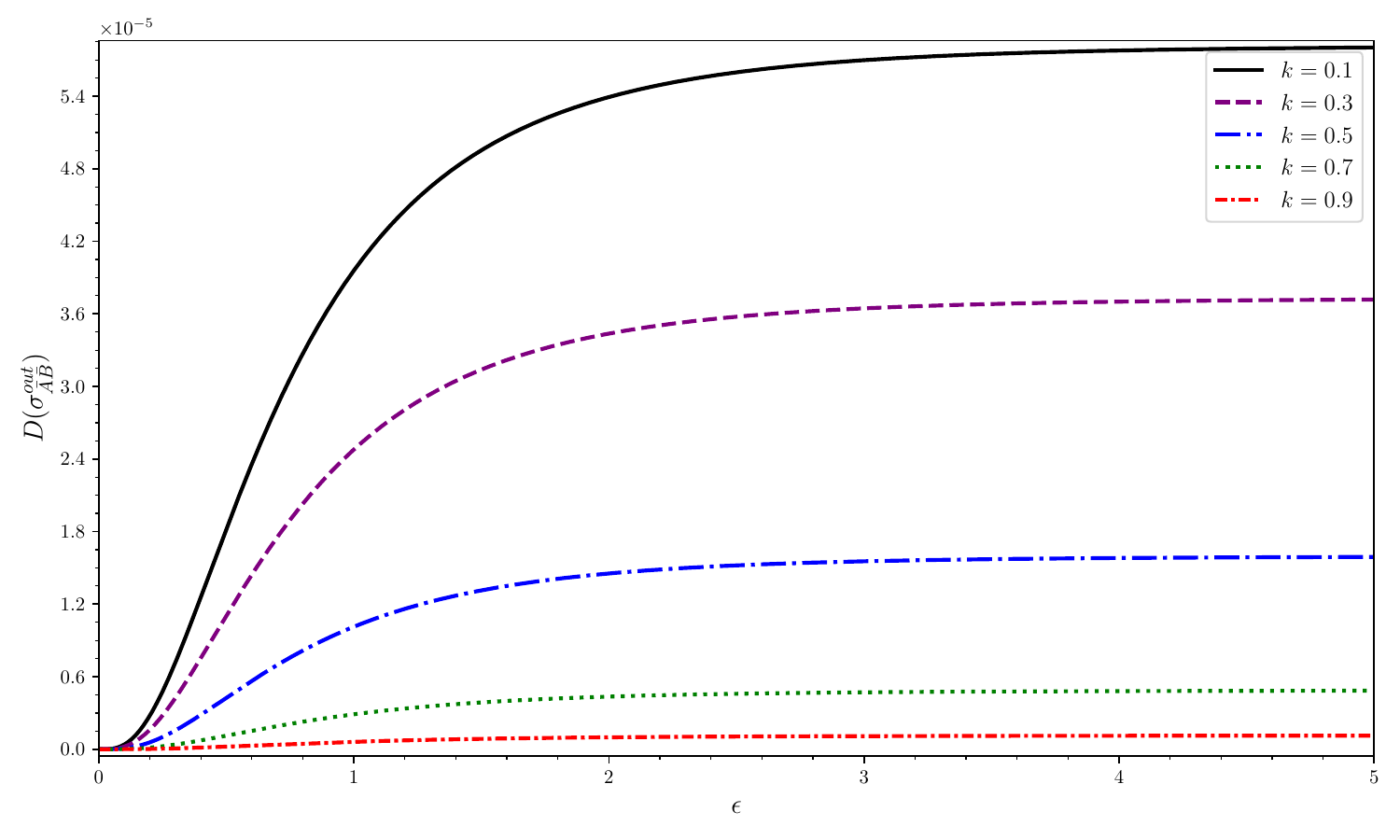}
		\\[-0.5em]
		\hspace*{1.2em}(a)
	\end{minipage}
	\hfill
	\begin{minipage}[b]{0.32\textwidth}
		\centering
		\includegraphics[width=\linewidth]{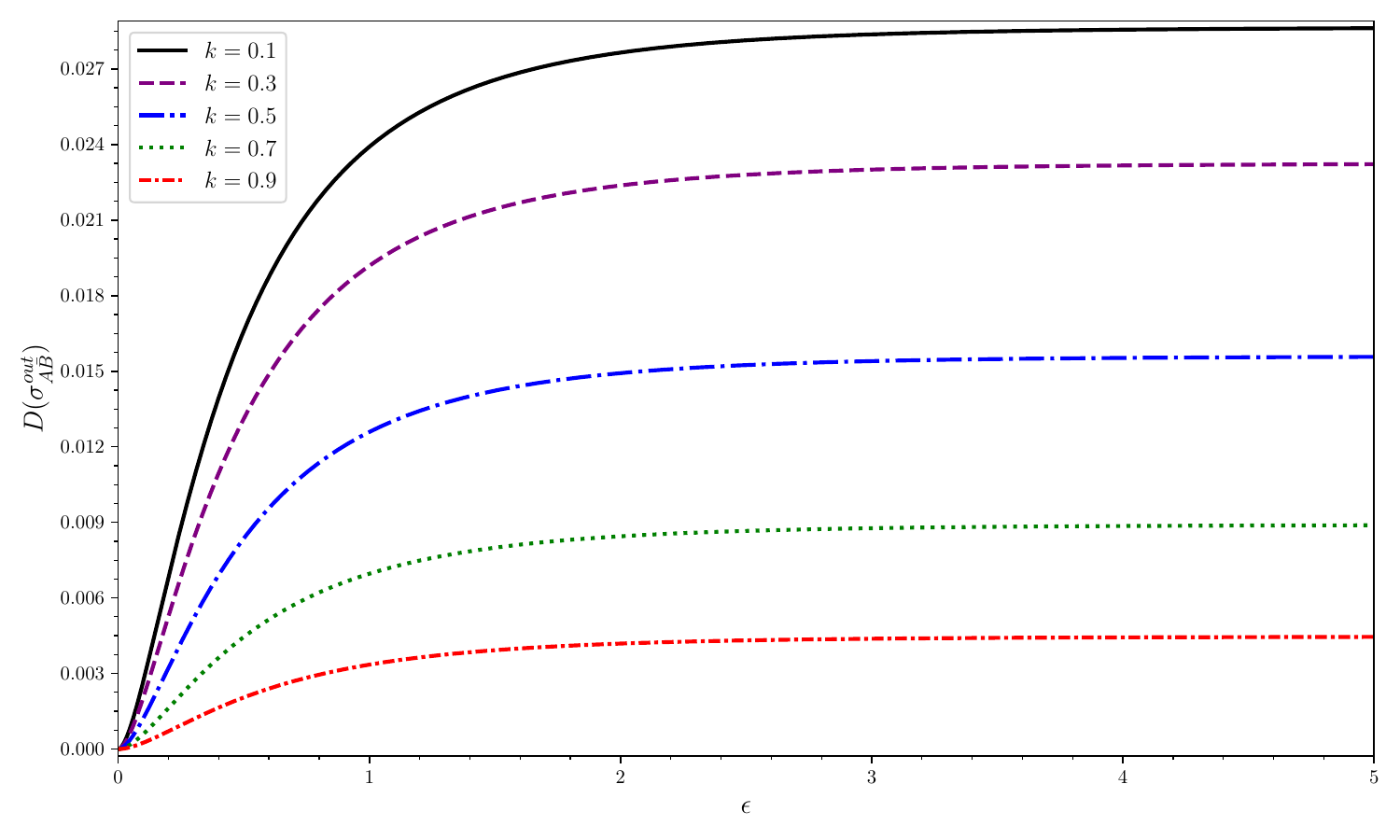}
		\\[-0.5em]
		\hspace*{1.5em}(b)
	\end{minipage}
	\hfill
	\begin{minipage}[b]{0.32\textwidth}
		\centering
		\includegraphics[width=\linewidth]{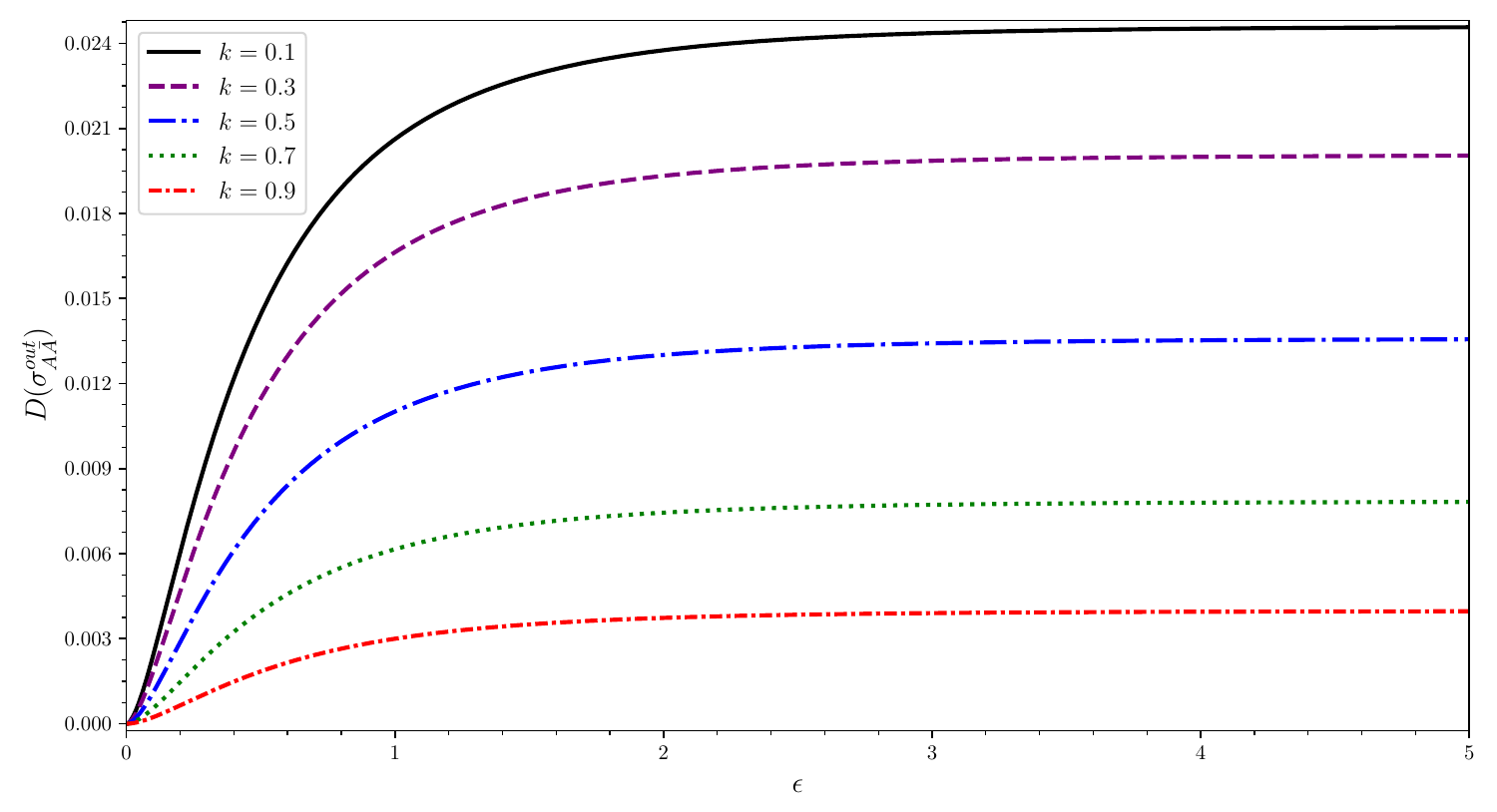}
		\\[-0.5em]
		\hspace*{1.5em}(c)
	\end{minipage}
	\caption{Panels (a), (b), and (c) show the quantum discord $D(\sigma^\mathrm{out}_{\bar{A}\bar{B}})$, $D(\sigma^\mathrm{out}_{A\bar{B}})$, and $D(\sigma^\mathrm{out}_{A\bar{A}})$ for the $\bar{A}$-$\bar{B}$, $A$-$\bar{B}$, and $A$-$\bar{A}$ mode pairs, respectively, as a function of the expansion volume $\epsilon$ for different momenta $k$, with fixed parameters $m=\upsilon=s=1$.}
	\label{fig:5}
\end{figure*}

In Figs.~\ref{fig:4} and \ref{fig:5}, we plot the quantum discord $D(\sigma^\mathrm{out}_{\bar{A}\bar{B}})$, $D(\sigma^\mathrm{out}_{A\bar{B}})$, and $D(\sigma^\mathrm{out}_{A\bar{A}})$ for the $\bar{A}$-$\bar{B}$, $A$-$\bar{B}$, and $A$-$\bar{A}$ mode pairs, respectively, as functions of the expansion rate $\upsilon$ (Fig.~\ref{fig:4}) and the expansion volume $\epsilon$ (Fig.~\ref{fig:5}) for different momenta $k$. We observe the following features:
(i) Cosmic expansion induces quantum discord between anti-Alice and anti-Bob, between Alice and anti-Bob, and between Alice and anti-Alice. Notably, Li et al\cite{Li2023Quantum}. demonstrated that cosmic expansion cannot generate quantum entanglement between anti-Alice and anti-Bob (or Alice and anti-Bob). Therefore, we argue that cosmic expansion gives rise to a form of non-entangled quantum correlation between these mode pairs;
(ii) The quantum discord for all these mode pairs increases with both the expansion rate $\upsilon$ and the expansion volume $\epsilon$, and decreases with increasing momentum $k$;
(iii) In the large-$\upsilon$ and large-$\epsilon$ limit, the quantum discord approaches an asymptotic value determined by the momentum $k$;
(iv) The quantum discord induced between Alice and anti-Bob is the largest, followed by that between Alice and anti-Alice, while that between anti-Alice and anti-Bob is the smallest; (v) Particles with smaller momentum $k$ help generate more quantum discord via cosmic expansion.

\begin{figure*}
	\centering
	\begin{minipage}[b]{0.32\textwidth}
		\centering
		\includegraphics[width=\linewidth]{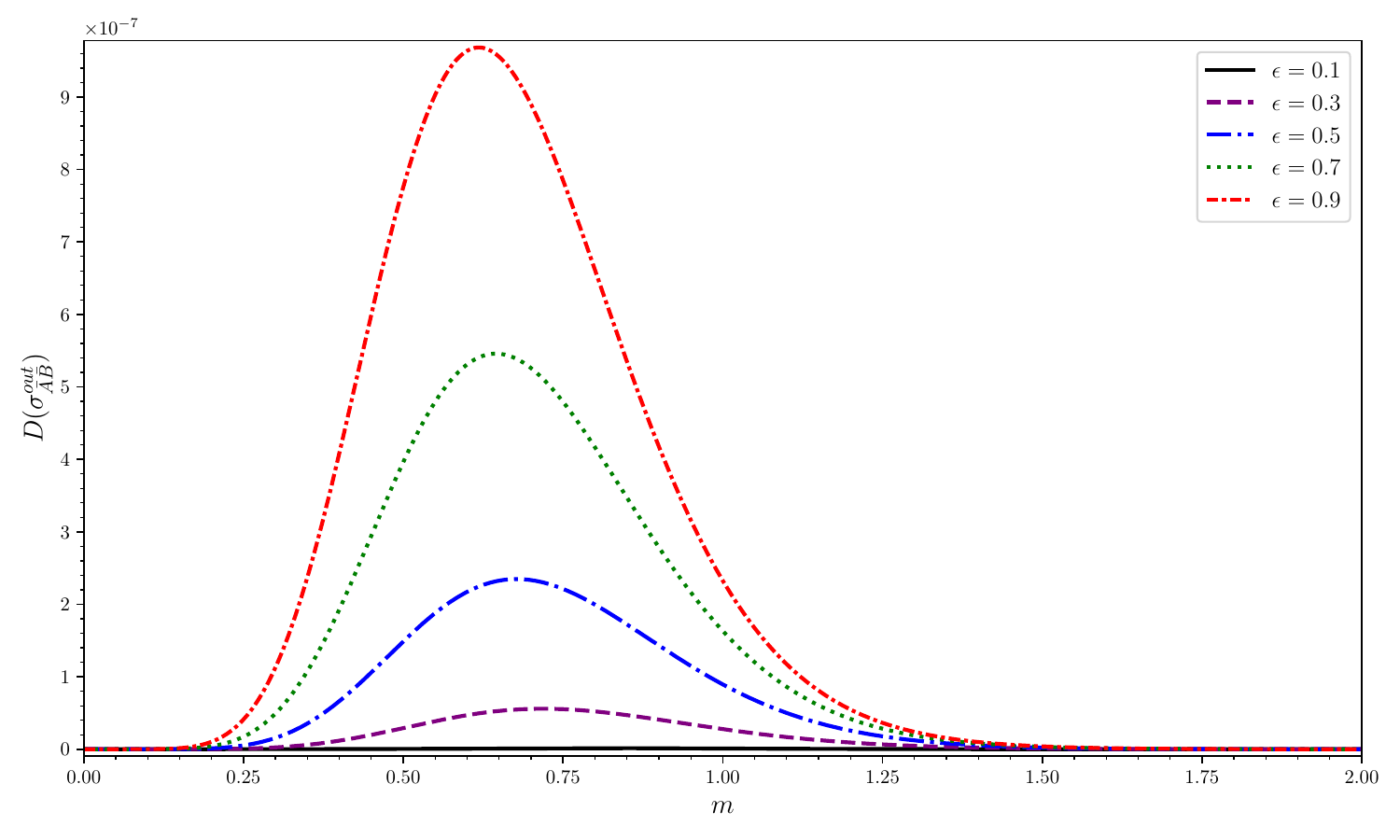}
		\\[-0.5em]
		\hspace*{0.7em}(a)
	\end{minipage}
	\hfill
	\begin{minipage}[b]{0.32\textwidth}
		\centering
		\includegraphics[width=\linewidth]{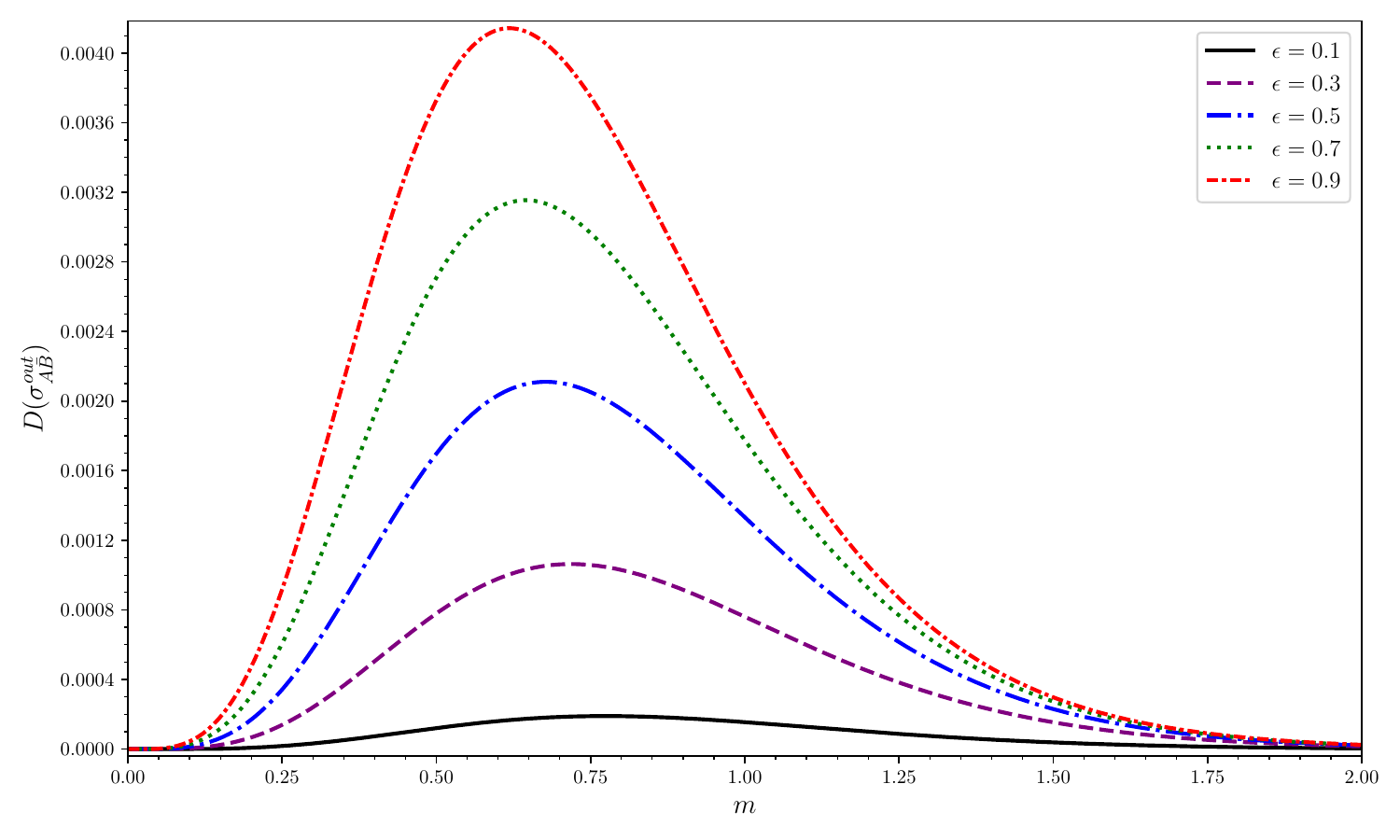}
		\\[-0.5em]
		\hspace*{1.4em}(b)
	\end{minipage}
	\hfill
	\begin{minipage}[b]{0.32\textwidth}
		\centering
		\includegraphics[width=\linewidth]{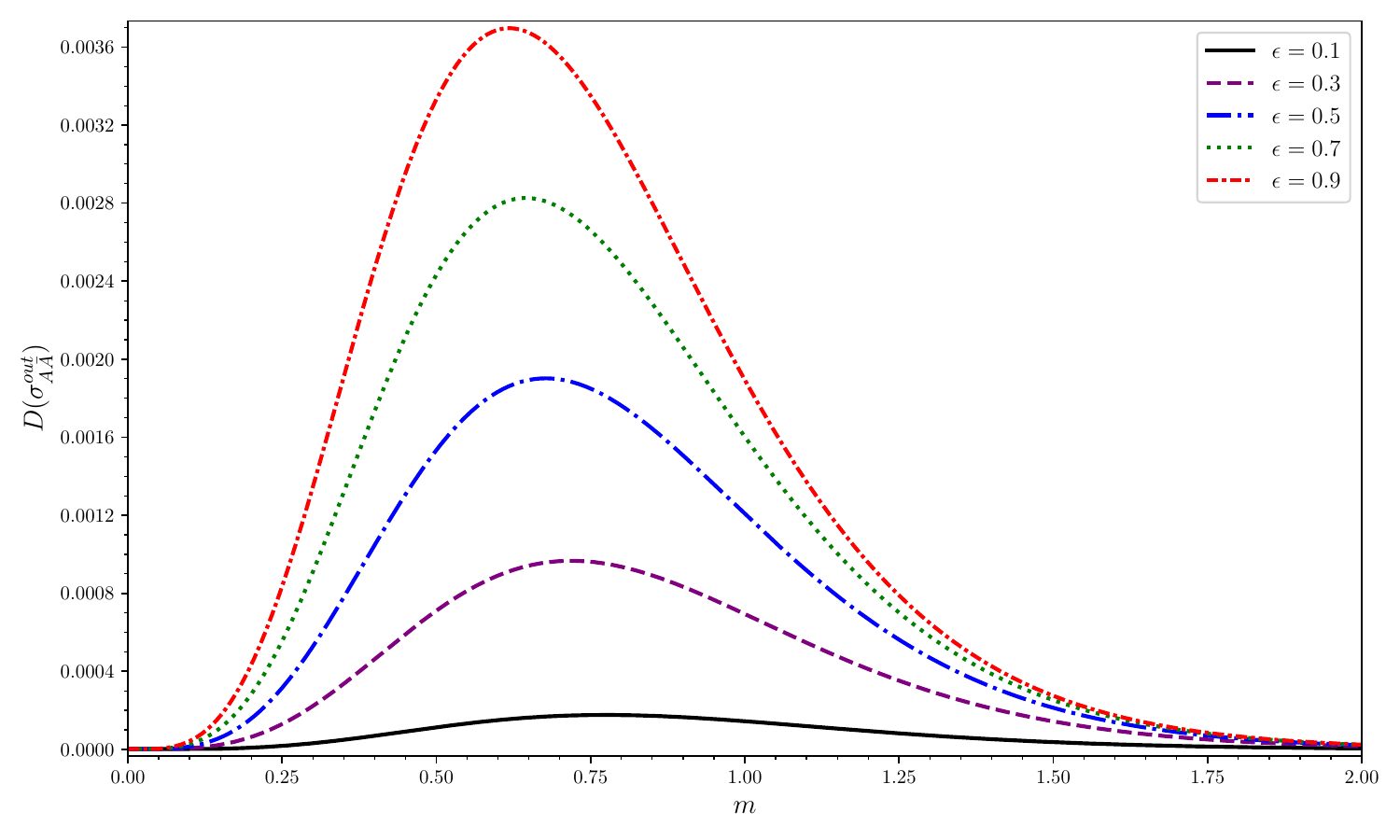}
		\\[-0.5em]
		\hspace*{1.4em}(c)
	\end{minipage}
	\caption{Panels (a), (b), and (c) show the quantum discord $D(\sigma^\mathrm{out}_{\bar{A}\bar{B}})$, $D(\sigma^\mathrm{out}_{A\bar{B}})$, and $D(\sigma^\mathrm{out}_{A\bar{A}})$ for the $\bar{A}$-$\bar{B}$, $A$-$\bar{B}$, and $A$-$\bar{A}$ mode pairs, respectively, as a function of the mass $m$  for different expansion volumes $\epsilon$, with fixed parameters $k=\upsilon=s=1$.}
	\label{fig:6}
\end{figure*}

In Fig.~\ref{fig:6}, we plot the quantum discord $D(\sigma^\mathrm{out}_{\bar{A}\bar{B}})$, $D(\sigma^\mathrm{out}_{A\bar{B}})$, and $D(\sigma^\mathrm{out}_{A\bar{A}})$ for the $\bar{A}$-$\bar{B}$, $A$-$\bar{B}$, and $A$-$\bar{A}$ mode pairs, respectively, as functions of the mass $m$ for different values of the expansion volume $\epsilon$. We observe the following features:
(i) For all these mode pairs, the quantum discord first increases from zero to an $\epsilon$-dependent maximum value with increasing mass $m$, and then decreases back to zero;
(ii) The quantum discord of bosonic fields with either very large or very small masses is insensitive to the expansion volume $\epsilon$, i.e., selecting particles with an appropriate mass facilitates the generation of more quantum discord via cosmic expansion.

\begin{figure*}
	\centering
	\begin{minipage}[b]{0.32\textwidth}
		\centering
		\includegraphics[width=\linewidth]{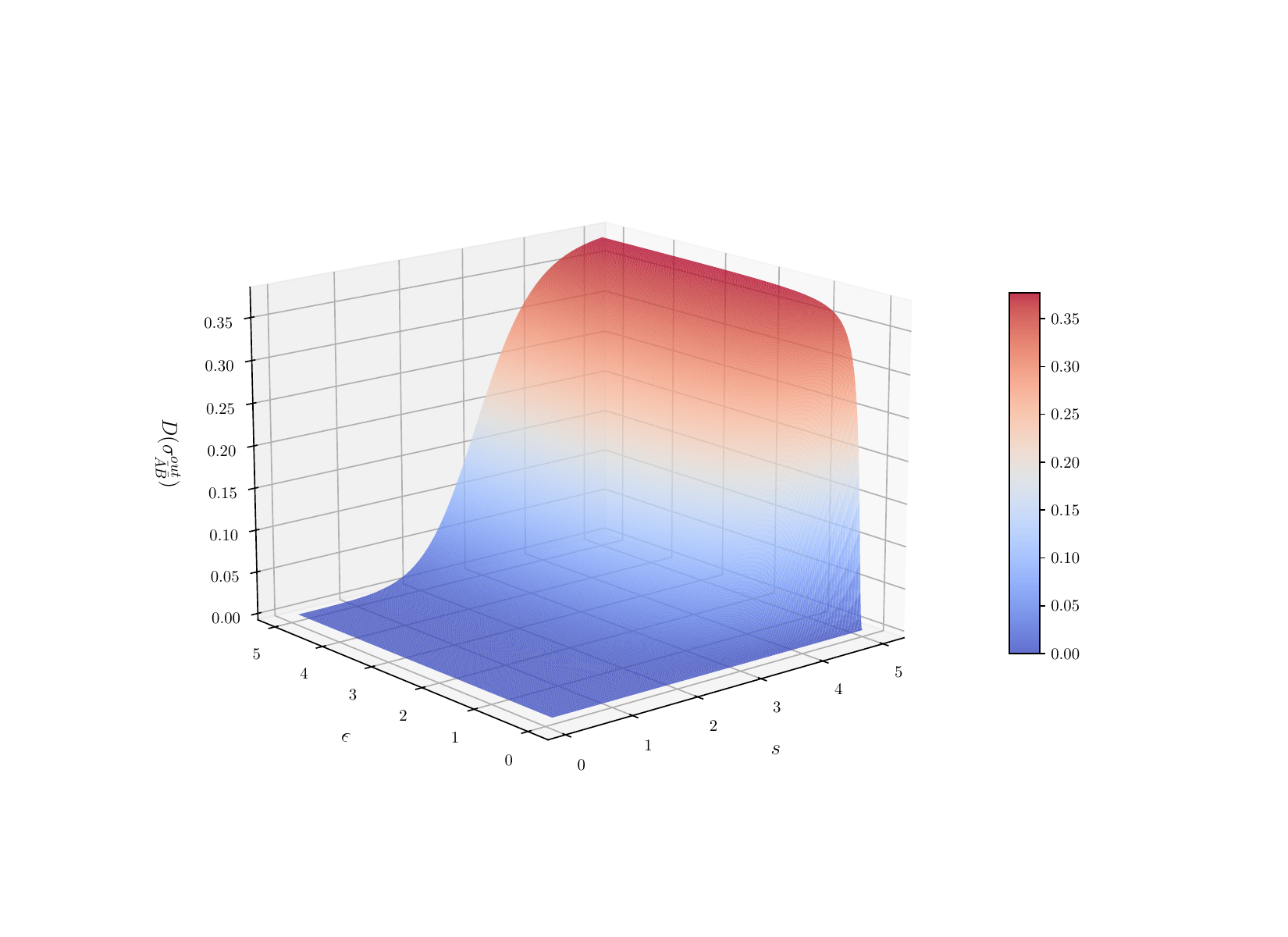}
		\\[-0.5em]
		\hspace*{0.7em}(a)
	\end{minipage}
	\hfill
	\begin{minipage}[b]{0.32\textwidth}
		\centering
		\includegraphics[width=\linewidth]{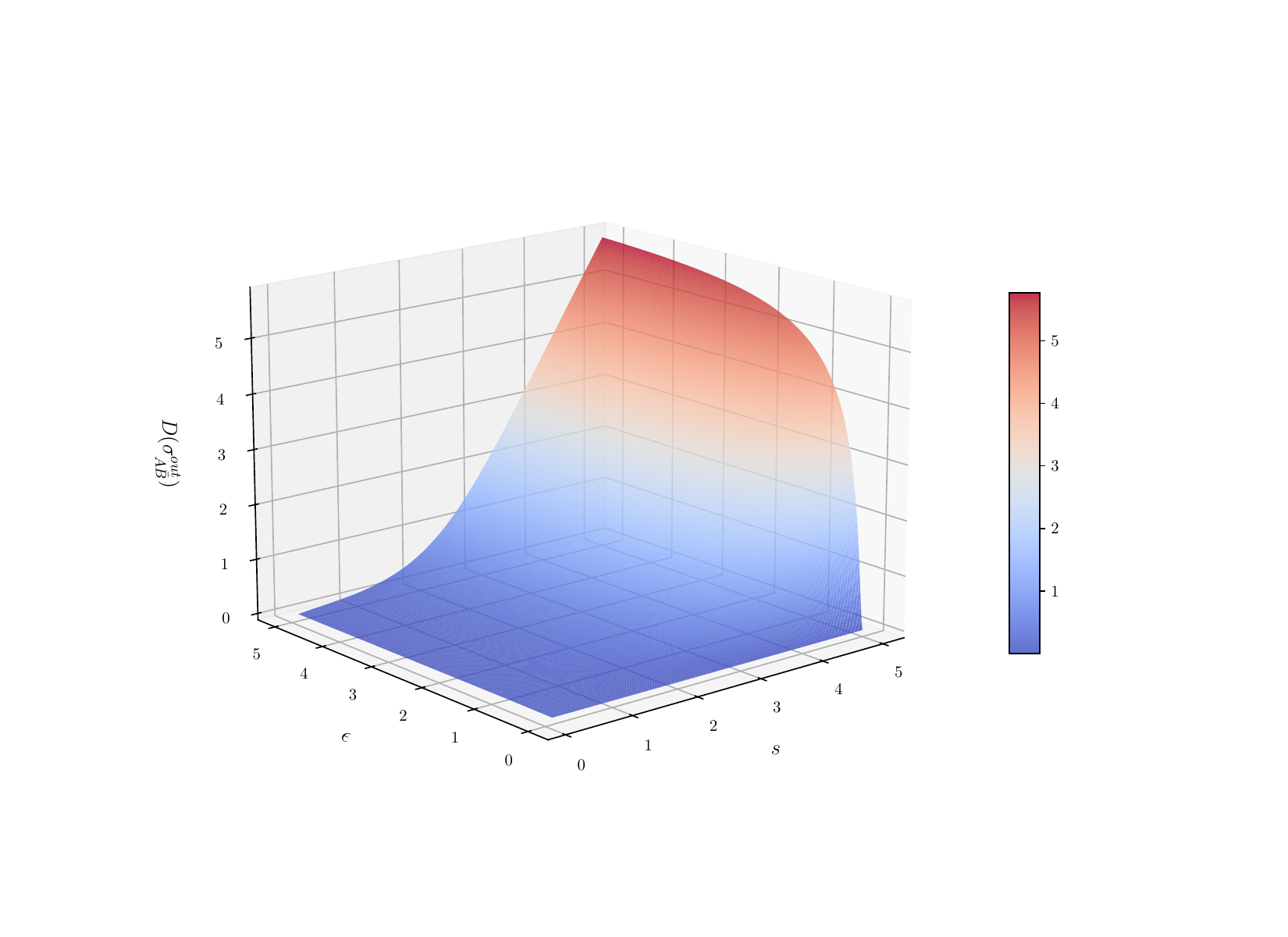}
		\\[-0.5em]
		\hspace*{1.4em}(b)
	\end{minipage}
	\hfill
	\begin{minipage}[b]{0.32\textwidth}
		\centering
		\includegraphics[width=\linewidth]{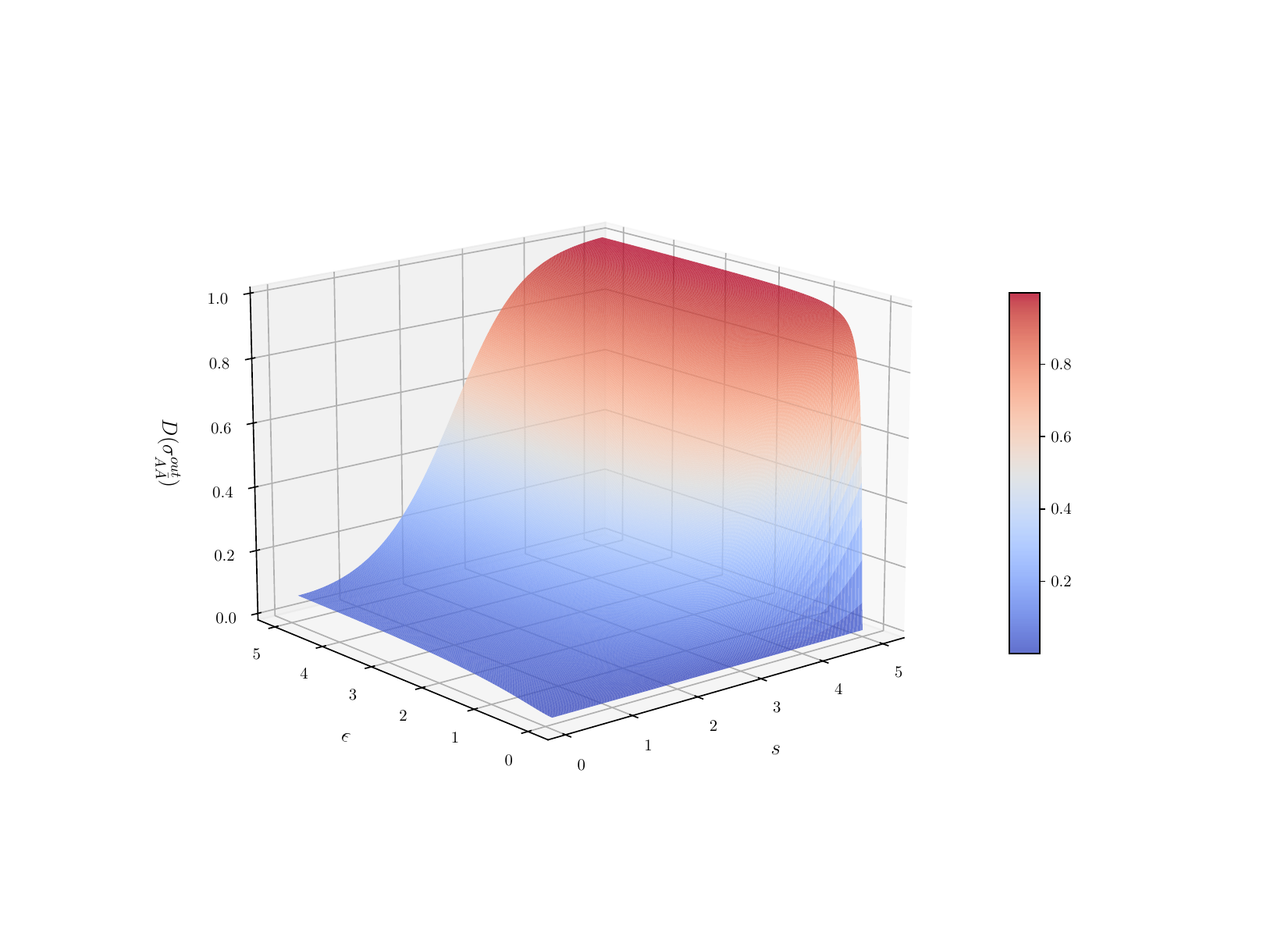}
		\\[-0.5em]
		\hspace*{1.4em}(c)
	\end{minipage}
	\caption{Panels (a), (b), and (c) show the quantum discord $D(\sigma^\mathrm{out}_{\bar{A}\bar{B}})$, $D(\sigma^\mathrm{out}_{A\bar{B}})$, and $D(\sigma^\mathrm{out}_{A\bar{A}})$ for the $\bar{A}$-$\bar{B}$, $A$-$\bar{B}$, and $A$-$\bar{A}$ mode pairs, respectively, as functions of the expansion volume $\epsilon$ and the squeezing parameter $s$, with fixed parameters $\frac{k}{2} = \frac{m}{2} = \upsilon = 1$.}
	\label{fig:7}
\end{figure*}

In Fig.~\ref{fig:7}, we plot the quantum discord $D(\sigma^\mathrm{out}_{\bar{A}\bar{B}})$, $D(\sigma^\mathrm{out}_{A\bar{B}})$, and $D(\sigma^\mathrm{out}_{A\bar{A}})$ for the $\bar{A}$-$\bar{B}$, $A$-$\bar{B}$, and $A$-$\bar{A}$ mode pairs, respectively, functions of the expansion volume $\epsilon$ and the squeezing parameter $s$. We observe the following features: (i) For all these mode pairs, the quantum discord rises with growing expansion volume $\epsilon$ under different squeezing parameters $s$, which agrees well with the behavior shown in Fig.~\ref{fig:5};
(ii) A larger squeezing parameter $s$ leads to a more pronounced enhancement of the induced quantum discord.

\begin{figure*}
	\centering
	\begin{minipage}[b]{0.32\textwidth}
		\centering
		\includegraphics[width=\linewidth]{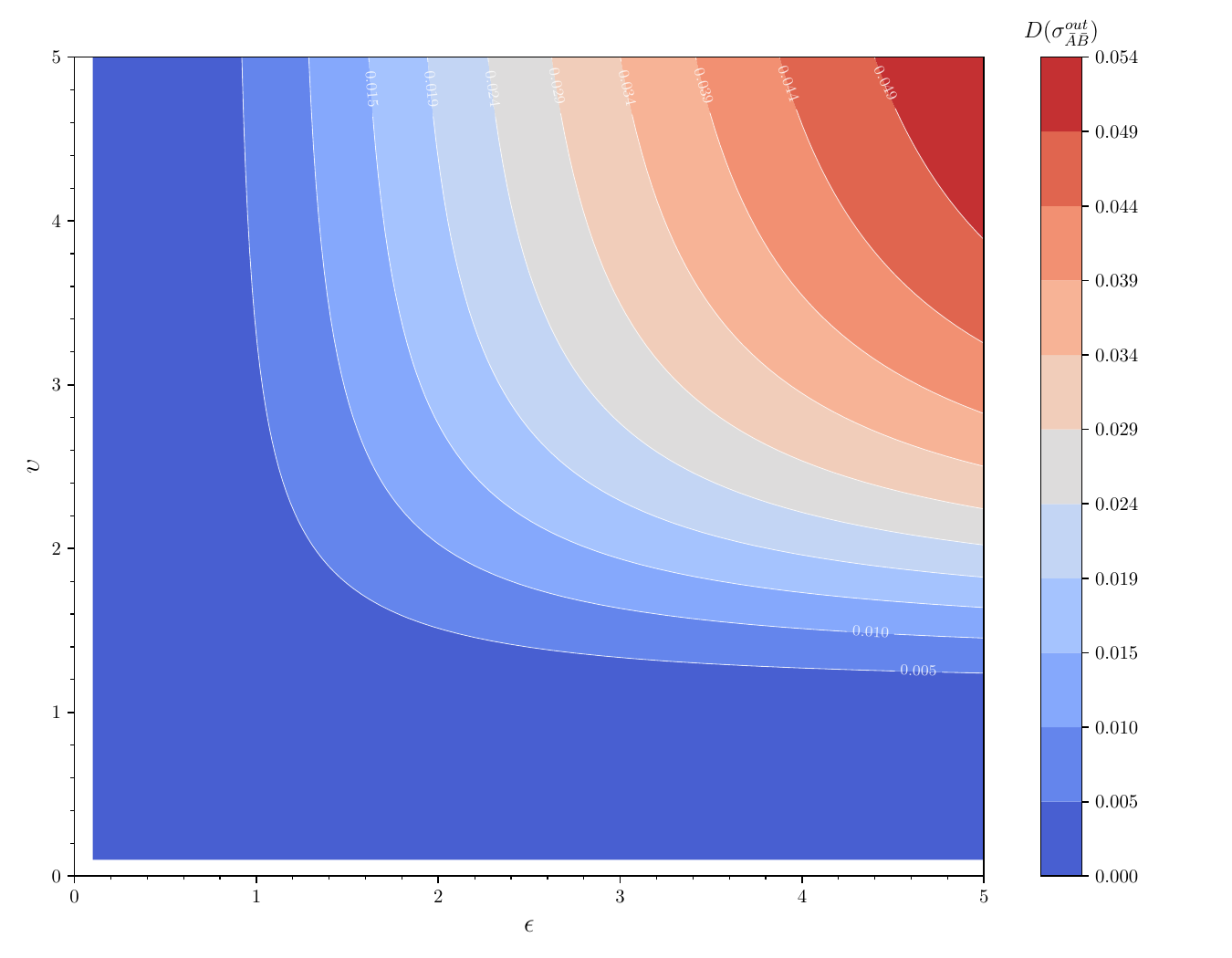}
		\\[-0.5em]
		\hspace*{-2.3em}(a)
	\end{minipage}
	\hfill
	\begin{minipage}[b]{0.32\textwidth}
		\centering
		\includegraphics[width=\linewidth]{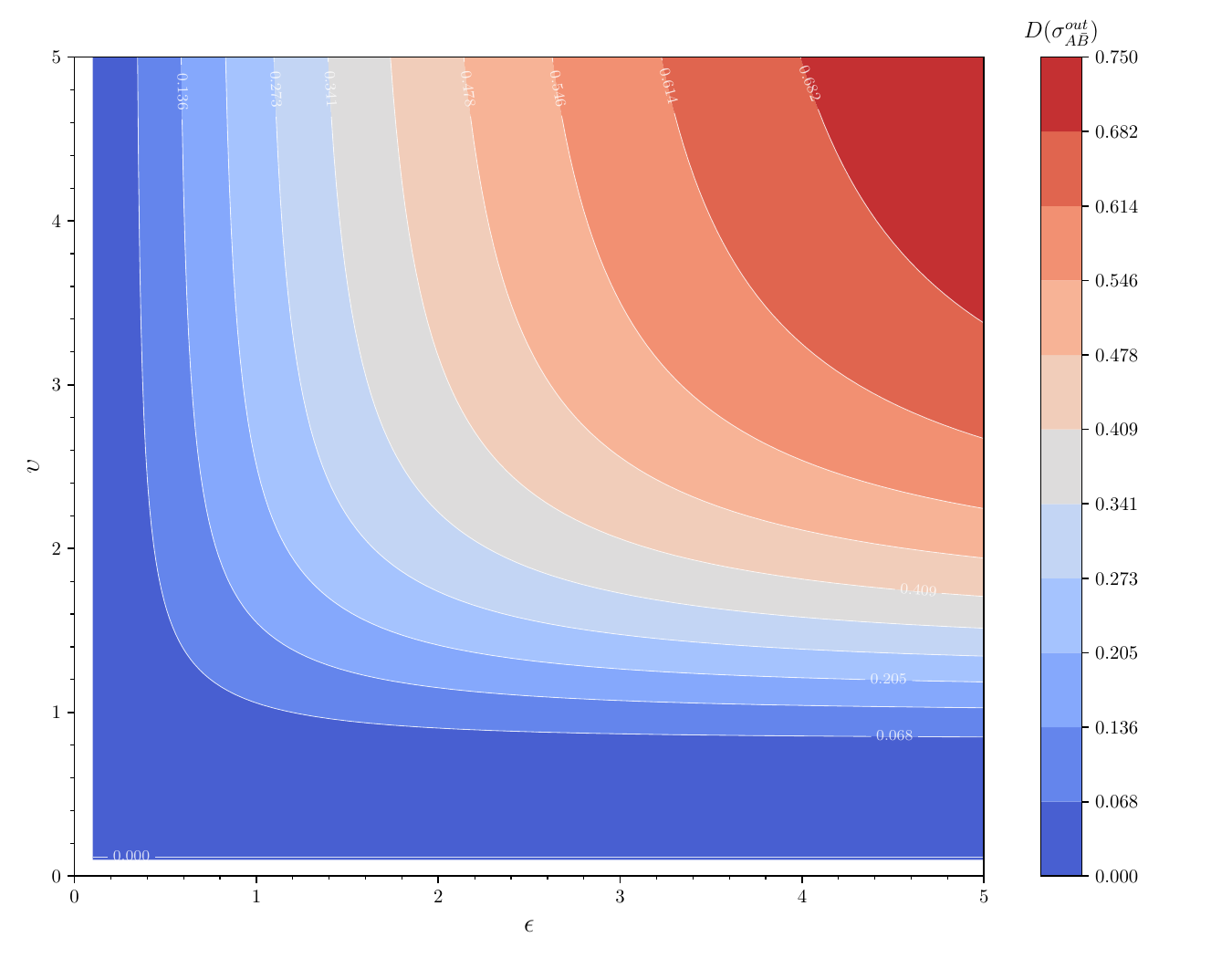}
		\\[-0.5em]
		\hspace*{-2.3em}(b)
	\end{minipage}
	\hfill
	\begin{minipage}[b]{0.32\textwidth}
		\centering
		\includegraphics[width=\linewidth]{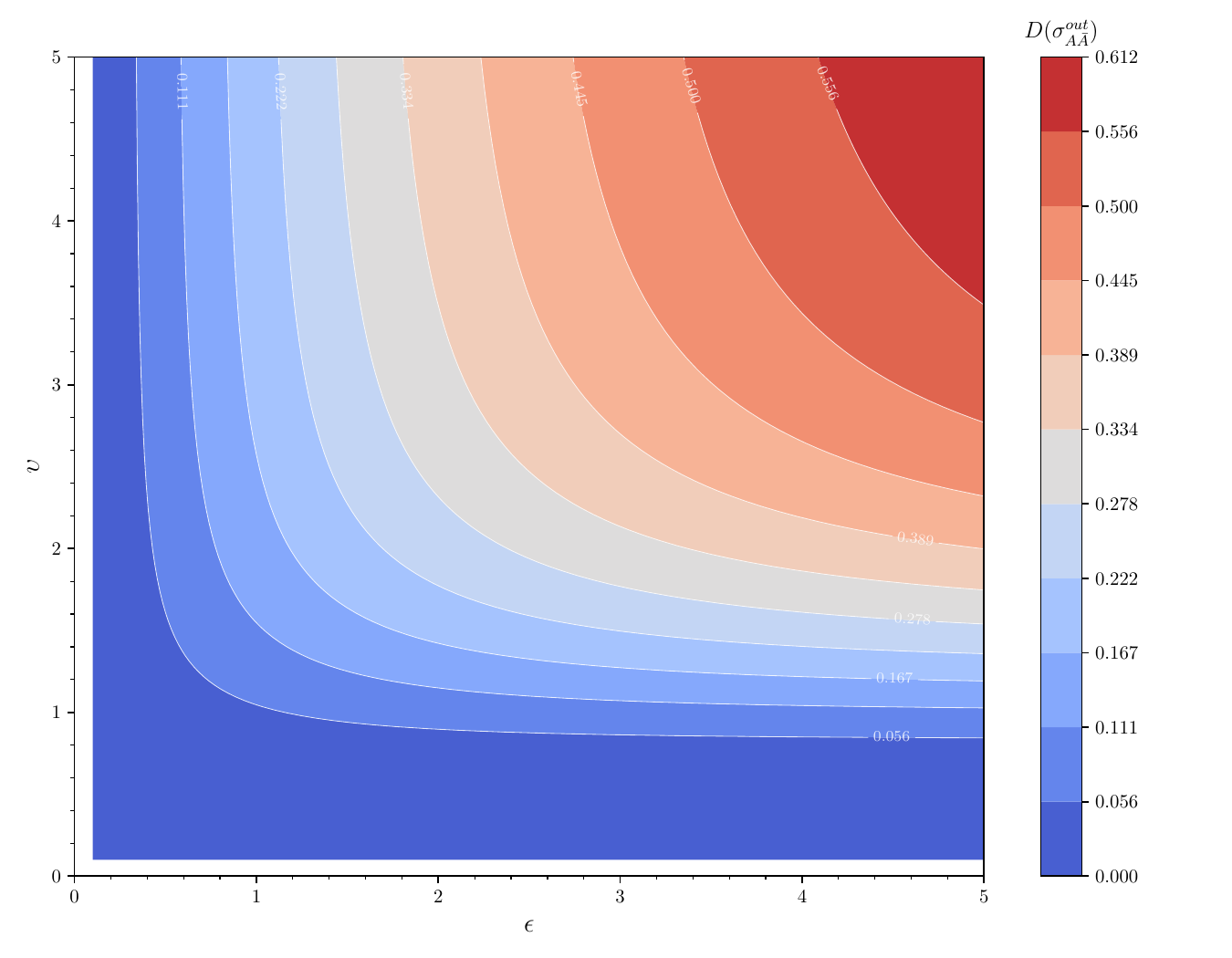}
		\\[-0.5em]
		\hspace*{-2.3em}(c)
	\end{minipage}
	\caption{Panels (a), (b), and (c) show the quantum discord $D(\sigma^\mathrm{out}_{\bar{A}\bar{B}})$, $D(\sigma^\mathrm{out}_{A\bar{B}})$, and $D(\sigma^\mathrm{out}_{A\bar{A}})$ for the $\bar{A}$-$\bar{B}$, $A$-$\bar{B}$, and $A$-$\bar{A}$ mode pairs, respectively, as functions of the expansion volume $\epsilon$ and the expansion rate $\upsilon$, with fixed parameters $\frac{k}{2} = \frac{m}{2} = s = 1$.}
	\label{fig:8}
\end{figure*}

In Fig.~\ref{fig:8}, we plot the quantum discord $D(\sigma^\mathrm{out}_{\bar{A}\bar{B}})$, $D(\sigma^\mathrm{out}_{A\bar{B}})$, and $D(\sigma^\mathrm{out}_{A\bar{A}})$ for the $\bar{A}$-$\bar{B}$, $A$-$\bar{B}$, and $A$-$\bar{A}$ mode pairs, respectively, functions of the expansion volume $\epsilon$ and the expansion rate $\upsilon$. Interestingly, the constraint behavior of the expansion rate $\upsilon$ and the expansion volume $\epsilon$ agrees well with that observed in Fig.~\ref{fig:3}. Furthermore, the quantum discord for all these mode pairs exhibits stronger sensitivity to the expansion rate $\upsilon$ than to the expansion volume $\epsilon$.

Based on the above analysis, we arrive at the following conclusions:
(i) As the expansion rate and expansion volume increase, the initial Gaussian quantum discord between Alice and Bob decreases, while the Gaussian quantum discord between anti-Alice and anti-Bob, between Alice and anti-Bob (or anti-Alice and Bob), and between Alice and anti-Alice (or Bob and anti-Bob) increases. This implies that cosmic expansion redistributes the initial quantum discord;
(ii) The quantum discord exhibits higher sensitivity to the expansion rate $\upsilon$ than to the expansion volume $\epsilon$;
(iii) By selecting particles with smaller momenta and appropriate masses, we can better extract information about the expanding universe and investigate the properties of quantum discord.

\section{conclusions}

In this work, we investigate the redistribution of continuous-variable quantum discord in an expanding spacetime. Our framework involves four modes: the bosonic mode $A$ observed by Alice, the bosonic mode $B$ observed by Bob, the antibosonic mode $\bar{A}$ observed by anti-Alice, and the antibosonic mode $\bar{B}$ observed by anti-Bob. We find that quantum discord exhibits stronger sensitivity to the expansion rate than to the expansion volume. We further show the redistribution of the initial quantum discord: the quantum discord between modes $A$ and $B$ decays with growing expansion rate and expansion volume, while the discord between modes $\bar{A}$ and $\bar{B}$, $A$ and $\bar{B}$ (or $\bar{A}$ and $B$), and $A$ and $\bar{A}$ (or $B$ and $\bar{B}$) can be induced by the expansion of the underlying spacetime. This implies that quantum discord is redistributed across these mode pairs. We also find that the induced quantum discord between modes $A$ and $\bar{B}$ (or $\bar{A}$ and $B$) is the largest, followed by that between modes $A$ and $\bar{A}$ (or $B$ and $\bar{B}$), whereas the quantum discord between modes $\bar{A}$ and $\bar{B}$ is the smallest. Through our analysis of quantum discord, we find that particles with smaller momentum and optimal mass constitute a more favorable candidate for extracting information regarding the expanding universe.

Our work substantially enriches the theoretical framework of quantum discord in expanding spacetime, and provides new perspectives as well as a solid theoretical foundation for further investigations.

\begin{acknowledgments}
This work is supported by the Natural Science Foundation of Hainan Province under Grant No. 125RC744; the China Scholarship Council (CSC).
\end{acknowledgments}

\bibliography{ref}

\end{document}